\documentclass[onecolumn]{aa} % for a paper on 1 column  
\usepackage{graphicx}
\usepackage{textcomp}                 %math symbols (\textmu)
\usepackage{txfonts}
\usepackage{amsmath}
\usepackage{natbib}
\usepackage[squaren,Gray]{SIunits}
\usepackage{color, colortbl}
\usepackage{array}
\usepackage{longtable,lscape}
\newcolumntype{L}[1]{>{\raggedright\let\newline\\\arraybackslash\hspace{0pt}}m{#1}}
\newcolumntype{C}[1]{>{\centering\let\newline\\\arraybackslash\hspace{0pt}}m{#1}}
\newcolumntype{R}[1]{>{\raggedleft\let\newline\\\arraybackslash\hspace{0pt}}m{#1}}

\newcommand{\Mearth}{$M_{\oplus}$}

\newcommand\Lp{$\text{L}^\prime$}

\definecolor{Gray}{gray}{0.9}

\begin{document}

   \title{New constraints on the dust surrounding HR\ 4796A}

   \author{
        J. Milli   \inst{1,2,3}
        \and D. Mawet \inst{3}
        \and C. Pinte \inst{1,2,4}
        \and    A.-M. Lagrange \inst{1,2} 
        \and D. Mouillet   \inst{1,2}
        \and  J. H. Girard \inst{3}
        \and J.-C. Augereau \inst{1,2}
        \and J. De Boer \inst{2,6}
        \and L. Pueyo \inst{5}
        \and \'E. Choquet \inst{5}
          }
          
   \institute{Universit\'e Grenoble Alpes, IPAG, F-38000 Grenoble, France \\
              \email{julien.milli@obs.ujf-grenoble.fr}
              \and 
              CNRS, IPAG, F-38000 Grenoble, France 
              \and
                European Southern Observatory (ESO), Alonso de C\'ordova 3107, Vitacura, Casilla 19001, Santiago, Chile
            \and
                UMI-FCA, CNRS/INSU France (UMI 3386) and Departamento de Astronom\'ia, Universidad de Chile,
Casilla 36-D Santiago, Chile    
            \and
                Space Telescope Science Institute, 3700 San Martin Drive, Baltimore MD 21218, USA
                \and
                Leiden Observatory, Leiden University, P.O. Box 9513, 2300 RA Leiden, The Netherlands 
             }

   \date{Submitted to Astronomy \& Astrophysics, March 31, 2014}
   \date{Received March 31, 2014; accepted February 9, 2015}

% \abstract{}{}{}{}{} 
% 5 {} token are mandatory
 
  \abstract
  % context heading (optional)
  % {} leave it empty if necessary  
   {HR\ 4796A is surrounded by a well-structured and very bright circumstellar disc shaped like an annulus with many interesting features: very sharp inner and outer edges, brightness asymmetries, centre offset, and suspected distortions in the ring.}
  % aims heading (mandatory)
   {We aim to constrain the properties of the dust surrounding the star HR\ 4796A, in particular the grain size and composition. We also want to confirm and refine the morphological parameters derived from previous scattered light observations, and reveal the dust spatial extent in regions unexplored so far due to their proximity to the star.}
  % methods heading (mandatory)
   {We  have obtained new images in polarised light of the binary system HR\ 4796A and B in the Ks and \Lp{} band with the NaCo instrument at the Very Large Telescope (VLT). In addition, we revisit two archival data sets obtained in the \Lp{} band with that same instrument and at \unit{2.2}{\micro\meter} with the NICMOS instrument on the Hubble Space Telescope. We analyse these observations with simulations using the radiative transfer code MCFOST to investigate the dust properties. We explore a grid of models with various dust compositions and sizes in a Bayesian approach.}
  % results heading (mandatory)
   {We detect the disc in polarised light in the Ks band and reveal for the first time the innermost regions down to $0.3\arcsec$ along the semi-minor axis. We measure a polarised fraction of $29\%\pm8\%$ in the two disc ansae, with a maximum occurring more than $13^\circ$ westwards from the ansae. A very pronounced brightness asymmetry between the north-west and south-east side is detected. This contradicts the asymmetry previously reported in all images of the disc in unpolarised light at wavelengths smaller than or equal to \unit{2.2}{\micro\meter} and is inconsistent with the predicted scattered light from spherical grains using the Mie theory. Our modelling suggests the north-west side is most likely inclined towards the Earth, contrary to previous conclusions. It shows that the dust is composed of porous aggregates larger than \unit{1}{\micro\meter}.}
      % conclusions heading (optional), leave it empty if necessary 
   {}


   \maketitle
%
%________________________________________________________________

\section{Introduction}

With a fractional luminosity of $5\times10^{-3}$, the disc around HR\ 4796A is one of the brightest cold debris disc systems among main-sequence stars. HR\ 4796A is an A0V star located at $72.8\pm1.7$pc. Together with HR\ 4796B, it forms a binary system with a projected separation of 560 AU. The dust was resolved at mid-infrared wavelengths \citep{Koerner1998,Jayawardhana1998}, at near-infrared wavelengths with NICMOS on the Hubble Space Telescope (HST) \citep{Schneider1999} and from the ground, with adaptive optics \citep[AO, ][]{Mouillet1997,Augereau1999,Thalmann2011,Lagrange2012_HR4796,Wahhaj2014}. The dust was also resolved at visible wavelengths with HST/ACS \cite{Schneider2009}.  The disc is confined to a narrow ring located at about 75AU, seen with a position angle (PA) of $26.8^\circ$ and inclined by $75.8^\circ$ with respect to pole-on. In the optical, the east side (both north and south) was seen brighter than the west side with a 99.6\% level of confidence \citep{Schneider2009}, which  led to the conclusion that the east side was inclined towards us with the common assumption of preferentially forward-scattering grains. Early modelling \citep{Augereau1999} showed that two components are needed to explain the spectral energy distribution (hereafter SED) up to \unit{850}{\micro\meter} and the resolved images up to thermal IR: a cold component, corresponding to the dust ring observed in scattered light, probably made of icy and porous amorphous silicate grains, plus a hotter one, with properties more similar to cometary grains, closer to the star. The need for the secondary component was debated by \citet{Li2003} but thermal images between $8$ and \unit{25}{\micro\meter} with improved spatial resolution confirmed the presence of hot dust within $\simeq$ 10 AU from the star \citep{Wahhaj2005}. More recent measurements of the far-infrared excess emission with APEX \citep{Nilsson2010}  and Herschel \citep{Riviere-Marichalar2013} confirmed the presence of the cold dust component and were able to constrain constrain its mass. The dust mass in grains below 1 mm was estimated at $0.146$\Mearth, using the flux density at \unit{870}{\micro\meter} by \citet{Nilsson2010}.

The high-angular resolution images of this system show that the disc ring is very narrow, with steep inner and outer edges, which were tentatively attributed to the truncation by unseen planets \citep{Lagrange2012_HR4796}. The morphology along the semi-minor axis  of the disc, at about $0.3\arcsec$, is poorly constrained, either because of the size of the coronagraph masking this inner region, as in \cite{Schneider2009} and \citet{Wahhaj2014}, or because of the star-subtraction algorithm, i.e., angular differential imaging \citep[ADI,][]{Milli2012}, which also removes much of the disc flux at such a short separation, as in \citet{Thalmann2011} and \citet{Lagrange2012_HR4796}. Polarimetric differential imaging (hereafter PDI) is a very efficient method for suppressing the stellar halo and revealing any scattered light down to $0.1\arcsec$. The technique PDI is based on the fact that the direct light from the star is unpolarised while the light scattered by the dust grains of the disc shows a linear polarisation. This technique was successfully used to characterise the protoplanetary discs around young stars (e.g. \citealt{Mulders2013,Avenhaus2014}). The first attempt to image the disc of HR\ 4796A using polarimetry was done by \citet{Hinkley2009}, who obtained a $6.5\sigma$ detection of the disc ansae at H band using the 3.6m AEOS telescope and a lower limit on the polarisation fraction of 29\%. Modelling the scattered light images both in intensity and polarisation is becoming a popular diagnostic tool for circumstellar discs because it brings constraints on the particles sizes and shapes (e.g. \citealt{Min2012,Graham2007}). This type of modelling breaks the degeneracies coming from modelling the SED alone. On HR\ 4796A, \citet{Debes2008} proposed a population of dust grains dominated by \unit{1.4}{\micro\metre} organic grains to explain the near-infrared dust reflectance spectra in intensity. In polarimetry, \citet{Hinkley2009} used simple morphological models, with an empirical scattering phase function and Rayleigh-like polarisability to conclude that their measurements are compatible with a micron-sized dust population. No attempts were made to use theoretical scattering phase function and to reconcile the SED-based modelling with the scattered light modelling. This will be discussed in this paper. 

In section \ref{sec_obs}, we present a new set of resolved images of the HR\ 4796A debris disc: a new detection in Ks polarised light, and a non-detection in \Lp{} polarised light. In section \ref{sec_past_observations}, we present two re-reductions of previously published images at those two wavelengths in unpolarised light. We analyse the morphology and the measured properties of the disc  in section \ref{sec_analysis}. An attempt to simultaneously fit the SED and scattered light images is presented in section \ref{sec_modelling} and discussed in section \ref{sec_discussion}.

%__________________________________________________________________

\section{New polarimetric observations}
\label{sec_obs}
\subsection{Presentation of the data}
The new observations are polarimetric measurements performed with VLT/NaCo in service mode in April and May 2013 \footnote{Based on observations made with ESO telescopes at the Paranal Observatory under Programme ID 091.C-0234(A)} \citep{Rousset2003,Lenzen2003}. Table \ref{tab_observing_log} provides a log of the observations. The star was observed in field tracking, cube mode\footnote{Individual frames are saved \citep{Girard2010}}, in the Ks (\unit{2.15}{\micro\metre}) and \Lp{} (\unit{3.8}{\micro\metre}) bands, with the S27 and L27 camera, respectively, providing a plate scale of 27mas/pixel. The polarimetric mode of NaCo uses a Wollaston prism to split the incoming light into an ordinary and an extraordinary beam, $I_{ord}$ and $I_{extra}$, separated by $3.30\arcsec$ and $2.97\arcsec$ in the Ks and \Lp{} band, respectively. A mask prevents the superimposition of the two beams but limits the field of view to stripes of $27\arcsec \times 3.3\arcsec$. A rotating half-wave plate (HWP) located upstream in the optical path enables the  selection of the polarisation plane. The polarisation of light can be represented by means of the Stokes parameters \citep[I,Q,U,V; ][]{Stokes1852}, where I is the total intensity, Q and U are the linearly polarised intensities, and V is the circularly polarised intensity. We did not consider circular polarisation  because NaCo does not have a quarter-wave plate to measure it. For each observation, we set the on-sky position angle to $45.5^\circ$  to align the two components HR\ 4796A and HR\ 4796B along the polarimetric mask so that both stars could be imaged on the same polarimetric stripe of the detector to enable cross-calibration. We used two dither positions to allow for sky subtraction, the two components being either centred on the bottom left or bottom right quadrant of the detector. One full polarimetric cycle consisted in four orientations of the HWP: $0^\circ$, $45^\circ$ to measure Stokes Q, and $22.5^\circ$, $67.5^\circ$ to measure Stokes U. We acquired two integrations (DIT $\times$ NDIT) with the same dither position per polarimetric cycle. 

\begin{table*}
\caption{Observing log}             % title of Table
\label{tab_observing_log}      % is used to refer this table in the text
\centering                          % used for centering table
\begin{tabular}{c c c c  c c c c c c }        % centered columns (4 columns)
\hline\hline                 % inserts double horizontal lines
Date & UT start/end & Filter & DIT\tablefootmark{a} & NDIT\tablefootmark{b} & $N_{exp}$\tablefootmark{c} & $N_p$\tablefootmark{d} & $t_{exp}$\tablefootmark{e} & $\tau_0$\tablefootmark{f} (ms) & Seeing ($\arcsec$) \\    % table heading 
\hline                        % inserts single horizontal line
   17-04-2013 & 02:24/02:50 & Ks & 0.35 & 110 & 2 &  3 & 231 & 8 & 0.8 \\      % inserting body of the table
   16-05-2013 & 03:03/04:07 & Ks & 0.5 & 100 & 2 & 8 & 800 & 6 & 1.4 \\      % inserting body of the table
   15-05-2013 & 02:49/03:56 & \Lp{} & 0.2 & 120 & 2 & 10 & 480  & 3 & 0.6 \\      % inserting body of the table
\hline                                   %inserts single line
\end{tabular}
\tablefoot{
\tablefoottext{a}{Detector Integration Time in seconds.}
\tablefoottext{b}{Number of DIT.}
\tablefoottext{c}{Number of exposures per polarimetric cycle.}
\tablefoottext{d}{Number of polarimetric cycles.}
\tablefoottext{e}{Total exposure time per HWP position in seconds.}
\tablefoottext{f}{Coherence time as indicated in the frame headers.}
}
\end{table*}

\subsection{Data reduction}

\subsubsection{Cosmetics and recentring}
All frames are sky-subtracted using the complementary dither positions and are then flat-field corrected. We noticed  that the cosmetized collapsed cubes of images are still affected by two different kinds of electronic noise in the Ks band. One of the sources of noise is described by \citet{Avenhaus2014} and affects some detector rows. It varies with a very short timescale within a data cube, it is therefore not corrected by a sky subtraction. We applied the row mean subtraction technique implemented by \citet{Avenhaus2014} in the H band to counteract this effect. The second source of noise only affects  the bottom right quadrant of NaCo and leaves a predictable square pattern only visible after the polarimetric subtraction. We ignored the frames where the star was located in that quadrant because this solution turned out to surpass any other filtering algorithm that we tried. The relevant stripes of the polarimetric mask were then extracted. Star centring was done by fitting a Moffat profile on the unsaturated wings of the star \citep{Lagrange2012}. We then performed a cross correlation between the ordinary and extraordinary images of the star to find the residual offset between those two frames and shifted the extraordinary image to the centre of the ordinary image. The uncertainty in the star centre is dominated by the  fit of the saturated PSF by a Moffat function. To evaluate it, we followed the methodology described in Appendix A.1 from \citet{Lagrange2012}, using the unsaturated PSF from the binary companion HR\ 4796B and found an error of $8.7 \text{mas}$ and $2.2 \text{mas}$ in the horizontal and vertical direction of the detector, respectively.

\subsubsection{Polarimetric subtraction}

\begin{figure}
   \centering
   \includegraphics[width=\hsize]{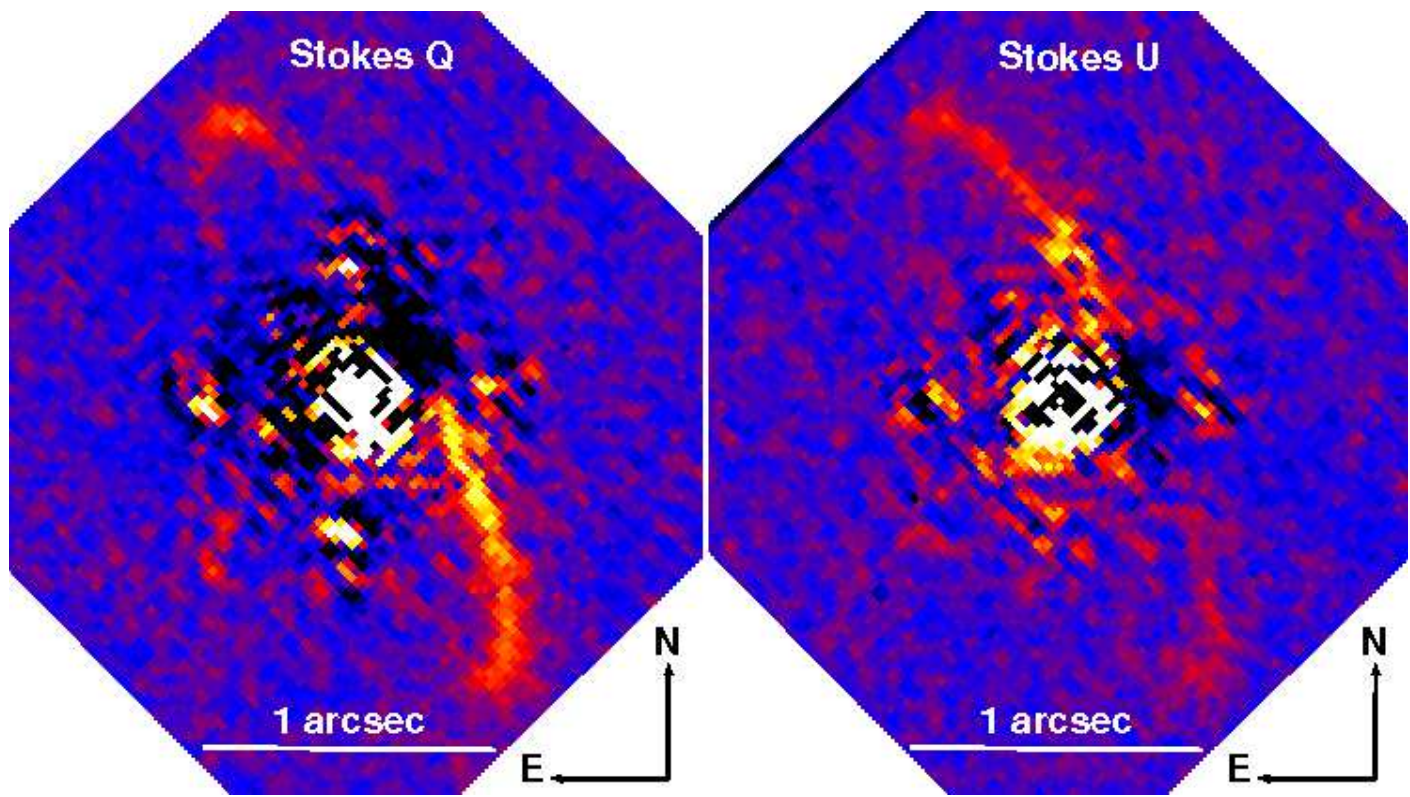}
      \caption{Stokes Q and U images in the Ks band before correcting for the instrumental polarisation.}                                     \label{fig_Ks_U_Q}
   \end{figure}

We briefly recall here the concepts of PDI in order to introduce the notations used hereafter. To derive the Stokes Q and U from the intensity measurements  $I_{ord}$ and $I_{extra}$, we performed the double ratio technique, following \citet{Tinbergen2005}. We also applied the double difference technique explained in detail in \citet{Canovas2011} but found in agreement with \citet{Avenhaus2014} that the former technique yielded a better signal-to-noise ratio (SNR). We computed the Q component of the degree of polarisation, $p_Q$, using 
\begin{equation}
p_Q = \frac{R_Q-1}{R_Q+1}
\end{equation}
 with 
\begin{equation}
R_Q=\sqrt{\frac{I_{ord}^{0^\circ}/I_{extra}^{0^\circ}}{I_{ord}^{45^\circ}/I_{extra}^{45^\circ}}}
.\end{equation}
 The Stokes Q parameter is then obtained with 
 \begin{equation}
Q = p_Q\times I_Q
,\end{equation}
 where $I_Q$ stands for the mean intensity in the images used with the HWP in position  $0^\circ$ and $45^\circ$: 
\begin{equation}
 I_Q = 0.5\times\left( I_{ord}^{0^\circ} + I_{extra}^{0^\circ} + I_{ord}^{45^\circ} + I_{extra}^{45^\circ} \right)
.\end{equation}
Replacing the upperscripts $0^\circ$ and $45^\circ$ by $22.5^\circ$ and $67.5^\circ$ yields the equations used to derive the Stokes U image. Both Stokes images are shown in Fig. \ref{fig_Ks_U_Q}. The radial and tangential Stokes parameters are then derived using
\begin{align}
& P_{\perp} = Q\text{cos}(2\phi)+U\text{sin}(2\phi) ,\\
& P_{\parallel} = -Q\text{sin}(2\phi)+U\text{cos}(2\phi), 
\end{align}
where $\phi$ is defined as the angle between the vertical on the detector and the line passing through the star located at $(x_0,y_0)$ and the position of interest  $(x,y)$ :
\begin{equation}
\label{eq_theta}
\phi = \text{arctan} \left( \frac{x-x_{0}}{y-y_{0}} \right) +\theta
.\end{equation}
The offset $\theta$ accounts for a slight misalignment of the HWP. It was estimated from the data, as explained in the next section. 
%The polarised intensity $pI$ is derived by $pI=\sqrt{U^2+Q^2}$, where p denotes the polarised fraction. 

\subsubsection{Correction for instrumental polarisation}
\label{sec_IR}

\begin{figure}
   \centering
   \includegraphics[width=\hsize]{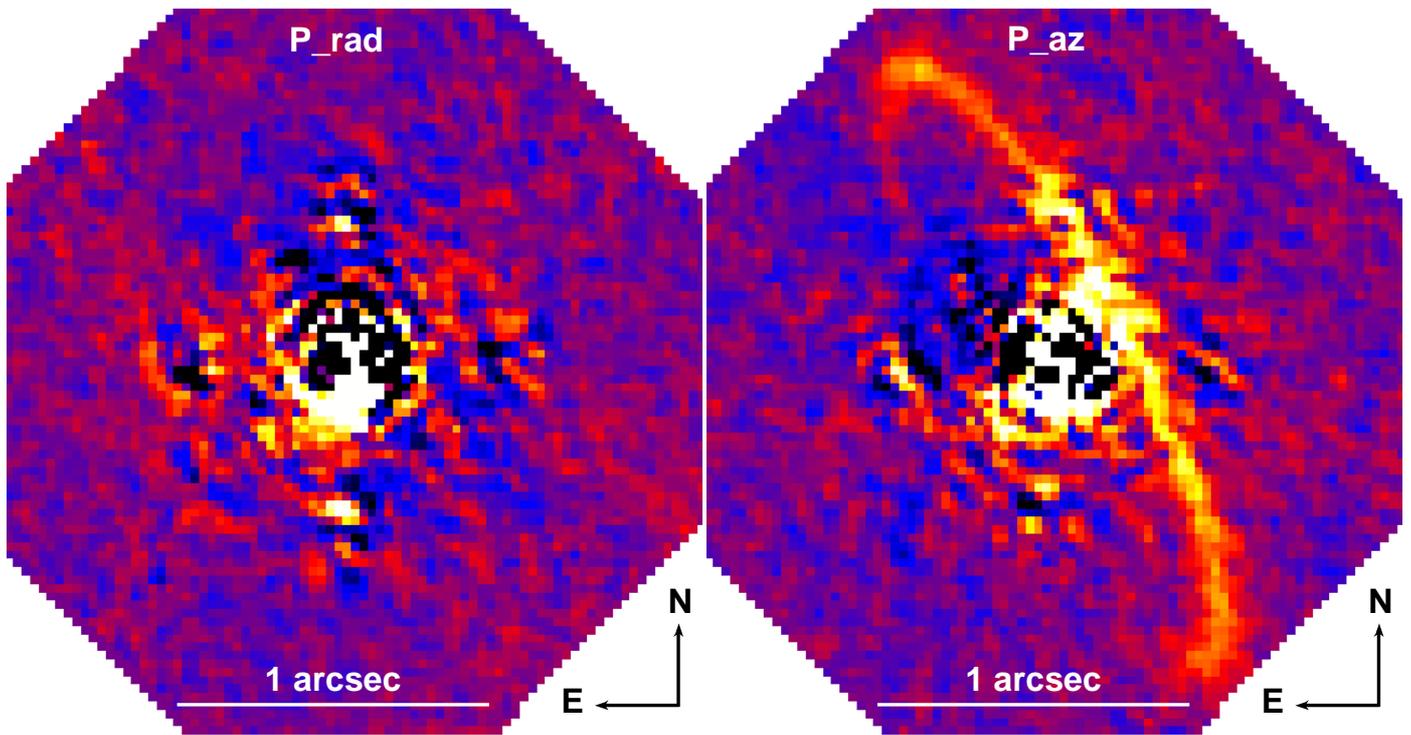}
      \caption{Radial and tangential Stokes parameters after correcting for the instrumental polarisation.}
         \label{fig_P_az_rad}
   \end{figure}

Because it stands at the Nasmyth focus of the VLT after the $45^\circ$ tilted mirror M3 and also has  many inclined surfaces, NaCo suffers from significant instrumental polarisation effects \citep{Witzel2011}.
We must correct for these effects to achieve the best polarimetric sensitivity and accuracy. Although HR\ 4796A is saturated, we use the unsaturated companion star to renormalize the images. We made the assumption that HR\ 4796B is not polarised and   that its flux in the ordinary and extraordinary image is equal. To do so, we re-normalized the ordinary and extraordinary image by the integrated flux of HR\ 4796B in the ordinary and extraordinary image.
We checked that the scaling factor was consistent with the value derived using the unsaturated halo of the star HR4796A, as done by \citet{Avenhaus2014}. Both values agree within $1\%$. 
Further instrumental polarisation remains, in particular we noticed that the disc flux in Stokes Q is unexpectedly higher than in Stokes U (Fig. \ref{fig_Ks_U_Q}). This effect represents a loss in polarimetric efficiency and is distinct from the former additive instrumental polarisation described above. This effect was also detected by \citet{Avenhaus2014} in the observation of HD\ 142527, and it is probably due to cross-talks between the U and V components of the Stokes vectors and to the lower throughput for Stokes U within NAOS (see the NAOS Mueller matrix in Eq. 16 of \citealt{Witzel2011}). They corrected for this effect  assuming that there must be as many disc pixels with $|Q| \geq |U|$ as with $|U| \geq |Q|$ and derived the Stokes U efficiency $e_{U}$ that satisfied this condition. While this assumption holds true for a pole-on disc like HD\ 142527, this is not necessarily true for an inclined disc such as HR\ 4796A, depending on its position angle on the detector. Instead, we assumed that the grain population is homogenous between the south-west (SW) and north-east (NE) ansae, therefore the polarisation fraction must be the same. Because the Stokes Q and U contribute differently to the polarised flux of each ansa, we were able to find a value for $e_{U}$ that yields the same polarisation fraction in both ansae averaged over the scattering angles between $88^{\circ}$ and $90^{\circ}$.  We derived $e_{U}=76\%$. This value is within the typical Stokes U efficiencies derived in the past with NaCo (from $55.0\%$ to $77.8\%$, \citealt{Avenhaus2014}). We also corrected for a possible misalignment of the HWP or a cross-talk between the Stokes Q and U. To do so, we introduced the offset $\theta$ in Eq. \ref{eq_theta}. For an optically thin disc, we expect from the Mie theory \citep{Mie1908} that the polarisation direction will be entirely tangential or, in some specific cases for large scattering angles, entirely radial. As we did not see any disc signal in the radial Stokes, we forced it to be 0 on average in the elliptical disc region. A value $\theta=-5.4^{\circ}$ was found to validate this property, in agreement with the typical values measured with NaCo (between $-3.7^{\circ}$ and $-7.0^{\circ}$, \citealt{Avenhaus2014}). The radial and tangential Stokes parameters, after correcting for the instrumental polarisation, are shown in Fig. \ref{fig_P_az_rad}. Most of the structures disappear in the radial polarisation.

\subsection{Subtraction of the point-spread function from intensity images}

In addition to the polarised intensity, the total intensity (Stokes I) contained in the sum $I_{ord}+I_{extra}$ is also  valuable information to characterise the light scattered by circumstellar matter. This sum contains the contribution of the star and the disc, it is therefore necessary to remove the contribution of the star to obtain the intensity of the disc. We used the binary component HR\ 4796B to build a library of reference point spread functions (PSF), that we later used to estimate the PSF to subtract from the intensity images of HR\ 4796A. To do so, we used the principal component analysis (hereafter PCA) technique described in \citet{Soummer2012}. Because HR\ 4796B is located at a projected separation of $7.7\arcsec$ where the distortion and variation in AO correction remains limited, and because it can be imaged simultaneously, this library was used as a representative set of reference PSF.

\subsection{Detection limits}

\subsubsection{Disc detection in the Ks band for the Stokes Q and U}

\begin{figure*}
   \centering
   \includegraphics[width=\hsize]{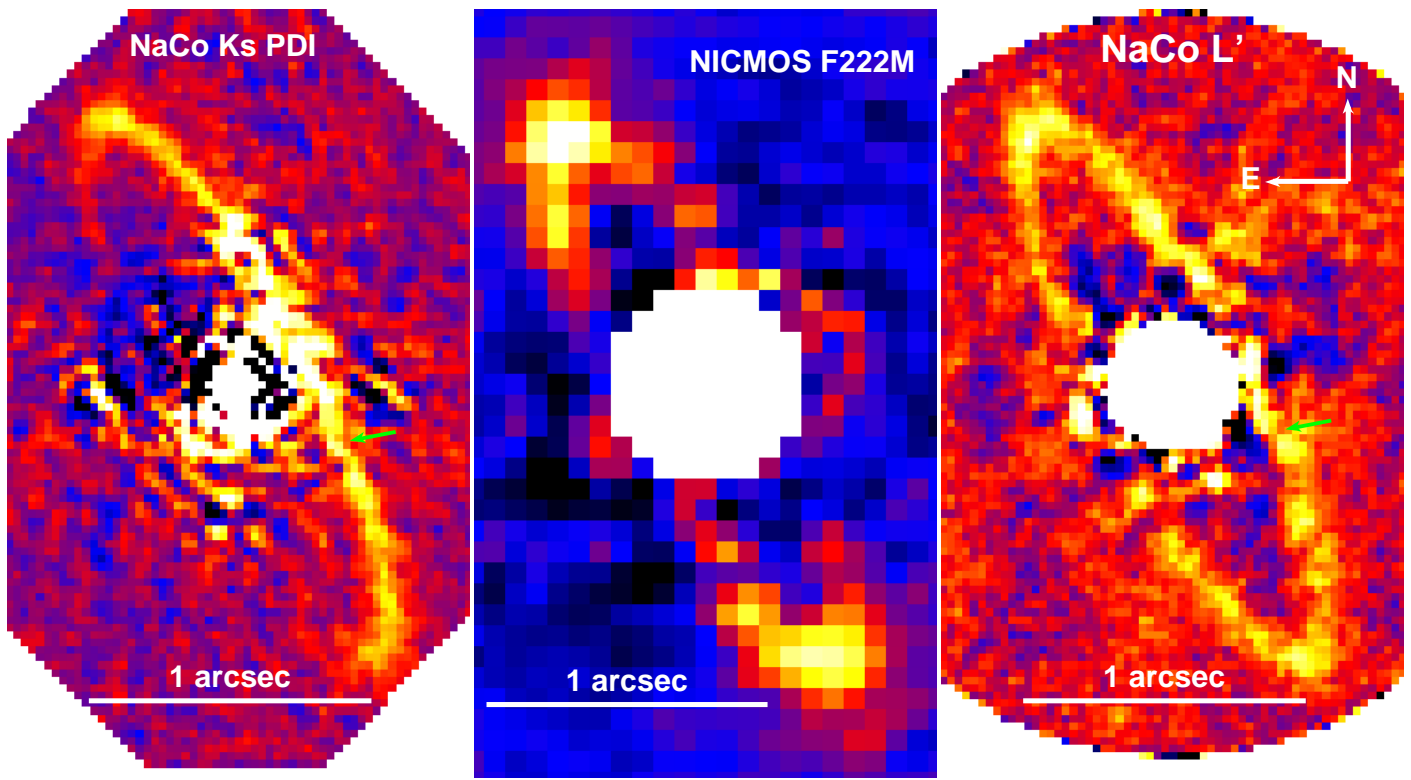}
      \caption{Left: Ks polarised intensity image of the disc. Middle: NICMOS image at \unit{2.2}{\micro\metre} (new reduction). Right: NaCo \Lp{} image (new reduction). The colour scale is linear.}
         \label{fig_Ks_F222_Lp}
   \end{figure*}

In the Ks band, we detect the disc in the Stokes Q and U images of both the April and May 2013 observations, but not in the Stokes I image. The May observations have a much higher integration time, and we did not find any improvement in combining both epochs. We therefore only present and analyse the images of this data set (Fig. \ref{fig_Ks_U_Q}, \ref{fig_P_az_rad} and \ref{fig_Ks_F222_Lp}, left). The disc is detected with a SNR of  9 in the ansae $1\arcsec$ away from the star ( $\sim5$, $\sim3$ for the Stokes Q and U image, respectively, with the Stokes U image having an increased noise level). The disc is detected down to $0.25\arcsec$ on the north-west (NW) side where it appears the brightest. The square pattern observed with a size of $0.76\arcsec$ is the waffle mode of NaCo and corresponds to a spatial frequency of $7\lambda/d$. 

\begin{figure}
   \centering
   \includegraphics[width=\hsize]{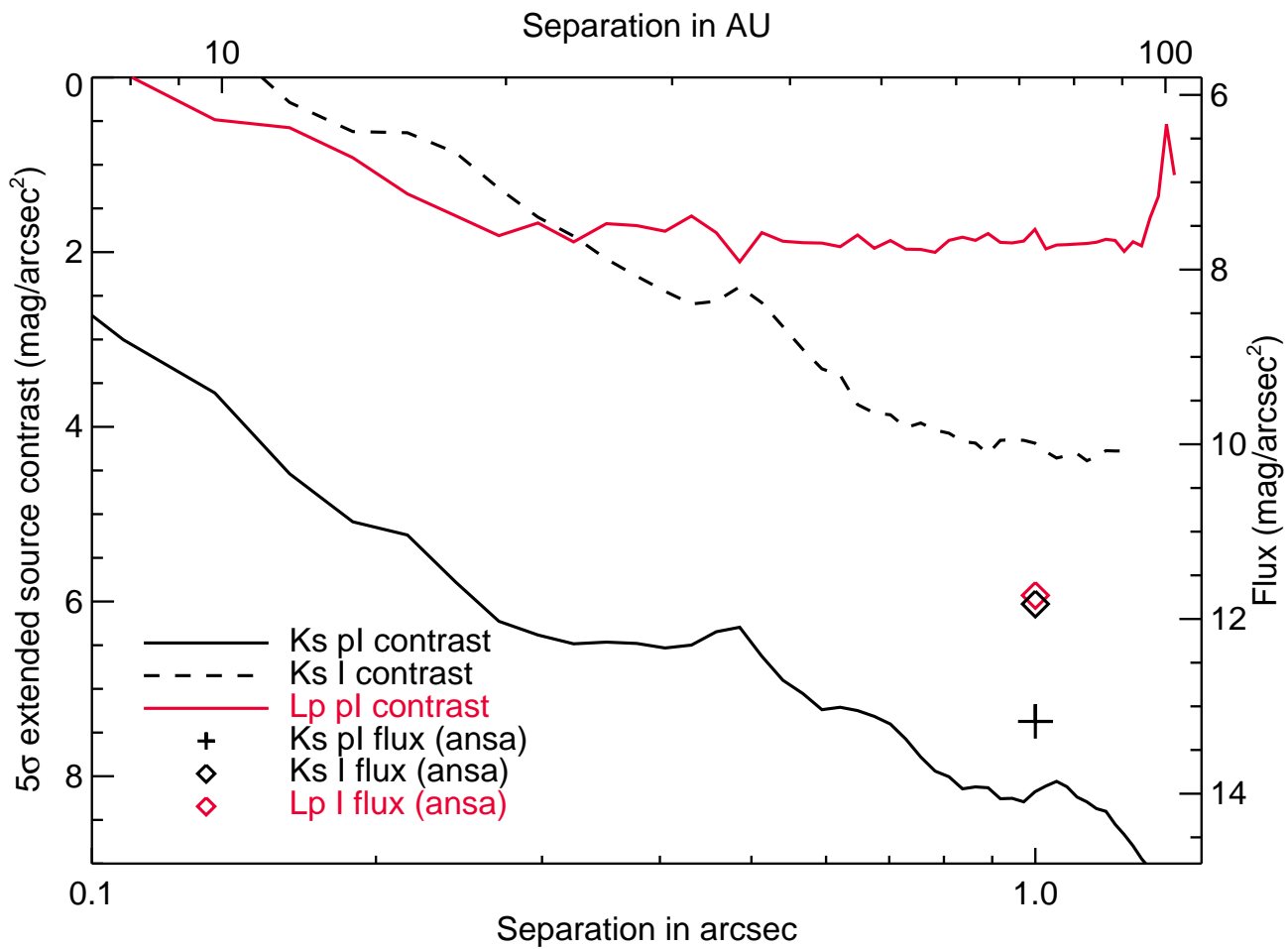}
      \caption{Contrast level reached in intensity (dotted curve) and polarised intensity (plain curves) for the Ks data (in black) and for the \Lp{} data (in red). The mean flux density in the ansae as measured with NaCo (black cross and red diamond) and NICMOS (black diamond) are overplotted.}
         \label{fig_det_lim_PDI_CI}
   \end{figure}

We used the unsaturated binary companion HR\ 4796B measured simultaneously to flux-calibrate our data. Using a K magnitude of 8.35 for the companion star, we are sensitive, in polarised light, to extended emission $2.5\text{mag/arcsec}^2$ fainter than the star at a separation of $0.1\arcsec$ and $8\text{mag/arcsec}^2$ $1\arcsec$ away from the star, as shown in Fig. \ref{fig_det_lim_PDI_CI}. We also overplotted the detected flux of the disc in polarised intensity.
Our contrast in total intensity is degraded by 4 magnitudes (black dotted curve). Despite using one of the most 
advanced PSF subtraction algorithms (PCA), we clearly cannot detect the disc in intensity given this level of contrast. Visual tests show that injecting a synthetic disc about 10 times brighter than the real disc leads to a $> 5\sigma$ detection in the ansae.

\subsubsection{Non-detection in the \Lp{} band}

While the disc was detected in the \Lp{} band in archival data (see section \ref{sec_naco_Lp_data}), we do not detect it in our new polarimetric data set either in polarimetry (Stokes Q or U)  or in intensity (Stokes I). The contrast level reached in the polarised intensity image is shown in Fig. \ref{fig_det_lim_PDI_CI} (red curve). In contrast to  the Ks band, the contrast curve is flat beyond $2\arcsec$, indicating that we are not limited by the residual light from the star. As  the data reduction procedure was done in the exact same way as in the Ks band, this poor result comes from  the shorter total integration time and the specificity of the \Lp{} band regarding the thermal emission and the optics transmission. The observational strategy implied a change in the dither position every seven min, which happened to be too slow in the \Lp{} band where the background is significantly higher and more variable than in the Ks band. Moreover, the Wollaston transmission drops to about 85\% due to an $MgF_2$ crystal absorption band at \unit{3.4}{\micro\metre}. 

From our photometry of the disc in the \Lp{} band (red diamond in Fig. \ref{fig_det_lim_PDI_CI}, see section \ref{sec_Lp_constraints} for details on the derivation), we conclude that our current polarisation detection limits are more than one magnitude above the disc flux density in the ansae. Therefore even if 100\% of the scattered light were polarised, we would still not detect it. 
 
\section{Analysis of archival data at \unit{2.2}{\micro\metre} and in the $\text{L}^\prime$ band }
\label{sec_past_observations}

\subsection{HST/NICMOS at \unit{2.2}{\micro\metre}}

Because the disc is not detected in Stokes I in our new polarimetric data, we used archival data of HR\ 4796A from HST/NICMOS in the F222M filter (central wavelength at \unit{2.22}{\micro\metre}). The star was observed at two roll angles on August, 12 2009 (Proposal id: 10167, PI: A. Weinberger) for a total integration time of 36min. We reprocessed the data using the reference star subtraction technique based on PCA with a library of NICMOS PSF \citep{Soummer2014}. The newly reduced image is shown in Fig.  \ref{fig_Ks_F222_Lp} (middle). Although these archival data were originally used to extract the photometry of the disc in \citet{Debes2008}, this image of the disc taken with the F222M filter was never published before.

\subsection{VLT/NaCo in the \Lp{} band}
\label{sec_naco_Lp_data}

We revisited previous observations of the disc in the \Lp{} band from \citet{Lagrange2012_HR4796} for three reasons:
\begin{enumerate}
\item to check whether our non-detection in polarisation is compatible with the existing detection of the disc in this band in intensity. In \citet{Lagrange2012_HR4796} we did not measure the photometry of the disc; 
\item to get additional constraints on the dust colour; and
\item to compare the morphological parameters of the disc using the same measurement method.
\end{enumerate}

Since the release of \citet{Lagrange2012_HR4796}, more efficient reduction algorithms that reveal faint extended emission from circumstellar discs were developed, e.g. PCA \citep{Soummer2012}. We re-reduced the data using this technique. The final image is shown in Fig.  \ref{fig_Ks_F222_Lp} (right), and has an increased SNR. 

\section{Analysis}
\label{sec_analysis}
\subsection{Two brightness asymmetries}

\begin{figure}
   \centering
   \includegraphics[width=\hsize]{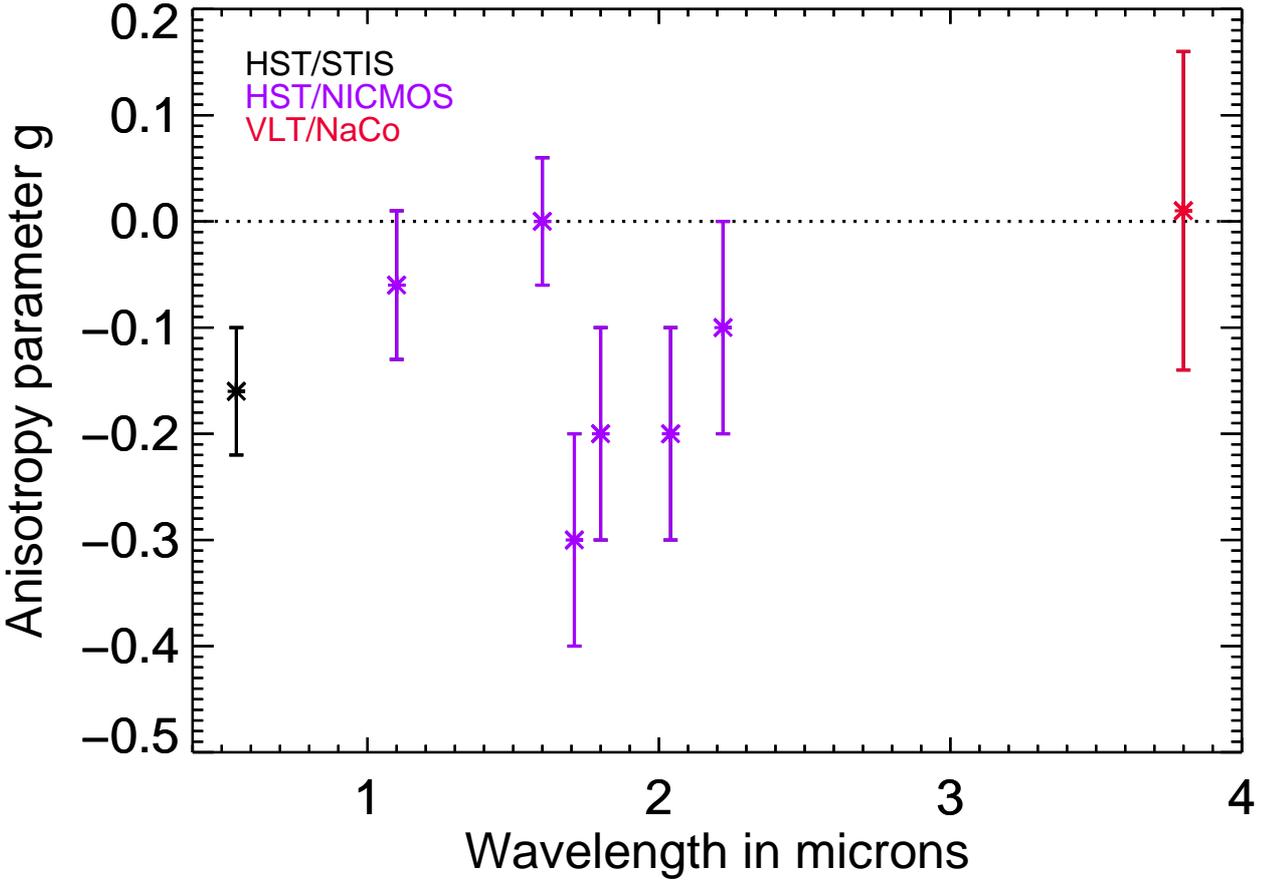}
      \caption{Evolution of the Henyey-Greenstein g coefficient with wavelength. It is based on the values derived by \citet{Debes2008} and  \citet{Schneider2009} for the HST images, using scattering angles between $5^{\circ}$ and $15^{\circ}$ from the major axis, and on our results for the VLT/NaCo new reduction at \Lp{} (section \ref{sec_Lp_constraints}). }
         \label{fig_g}
   \end{figure}

The most puzzling feature revealed by our new polarimetric observations is a very significant brightness asymmetry with respect to the semi-major axis of the star: the NW side is much brighter than the south-east (SE) side (Fig. \ref{fig_Ks_F222_Lp}  left). The first light images of the Gemini Planet Imager (GPI) on HR\ 4796A also detects this asymmetry \citep{Perrin2014}. The polarised flux reaches $3.4\text{mJy/arcsec}^{2}\pm0.4$ in the NE ansa whereas the SE side of the disc is not detected in polarimetry, its flux being below the noise level within $0.7\arcsec$.This new asymmetry contrasts with the NICMOS image (Fig. \ref{fig_Ks_F222_Lp}  middle) showing the opposite trend, and with all previous scattered light observations below \unit{3}{\micro\metre} \citep{Debes2008}. 
From these previous observations, and assuming that dust is generally preferentially forward-scattering, it was assumed that the SE side was  inclined towards the Earth \citep{Schneider2009}. To be consistent with our modelling presented in section \ref{sec_modelling}, we  use the opposite assumption as our baseline scenario: the NW side is the side inclined towards the Earth. Under this assumption, Fig. \ref{fig_g} summarises the anisotropic behaviour of the dust reported in intensity images in the optical and near-infrared. The anisotropy of scattering is generally described by the empirical Henyey-Greenstein phase function, which is parametrized by a single coefficient called g. This parameter between -1 and 1 is  zero for an isotropic disc, positive for a forward-scattering disc and negative in case of backward scattering. Therefore in our baseline scenario, a negative value of g indicates that the SE side is brighter than the NW side. Although there is a jump after \unit{1.6}{\micro\metre}, the overall trends seems to indicate that the dust  becomes more and more isotropic ($g \sim 0$) at longer near-infrared wavelengths. This trend is confirmed by our new reduction at $L^{\prime}$ (Fig. \ref{fig_Ks_F222_Lp} right), compatible with an isotropic disc (see section \ref{sec_Lp_constraints}). This interesting behaviour is  discussed in  section  \ref{sec_anistotrpy_scat_vs_lambda}.

A second brightness asymmetry was reported in the past with respect to the NE and SW ansae. This asymmetry is thought to be partially caused by pericentre glow due to the offset of the disc \citep{Schneider2009,Moerchen2011}, but this explanation is not sufficient to account for the amplitude of the asymmetry \citep{Wahhaj2014}. Our  new polarisation image also shows this asymmetry with a ratio between the NE and SW ansae of $1.07\pm0.15$ (with a $3\sigma$ uncertainty). In the unpolarised NICMOS image, this ratio is slightly less, $1.04\pm0.36$, but it reaches $1.34\pm0.23$ in the \Lp{} image. 

\subsection{Ring geometry}

\subsubsection{Constraints from polarised observations at Ks}

We note a small distortion in the SW at a separation of $0.4\arcsec$ and a position angle of $235^\circ$ in the Ks polarised image (green arrows in Fig. \ref{fig_Ks_F222_Lp}). This feature is also present at the same position in H band and  optical images \citep{Thalmann2011}, and in our \Lp{} image (Fig. \ref{fig_Ks_F222_Lp} right, already detected in the initial reduction by \citealt{Lagrange2012_HR4796}). 

To compare the morphology of the disc in polarised light with previous measurements, we interpret the disc as an inclined circular ring and measured its centre position with respect to the star, its radius, its inclination, and its position angle (PA). We build a scattered light model of the disc in intensity using the GRaTeR\footnote{GRaTeR is very fast and optimized for optically thin debris discs, and it is therefore more adapted to forward modelling than MCFOST.} code \citep{Augereau1999} with the same parametrization as presented in \citet{Lagrange2012_HR4796}. 
The model geometry is defined by six free parameters: the centre offset along the semi-major axis ($x_c$) and semi-minor axis ($y_c$), the inclination $i$, the PA, the radius $r_0,$ and a scaling factor to match the disc total flux.  We model the SE to NW asymmetry with a smooth function peaked on the NW side along the semi-minor axis and cancelling $30^\circ$ away from the ansa westwards. Our aim is not to constrain the dust properties by modelling the scattering phase function, rather  to build a map of the disc that is consistent with the disc morphology. Constraining the grain properties is the focus of  section $\ref{sec_discussion}$.

We then minimise a chi squared $\chi^2$ between our synthetic image and our model. The results are shown in Table \ref{tab_disc_param_Lp} (first row). The offset of the disc evaluated by \citet{Schneider2009} in the optical to $x_c=1.4\pm0.4$AU and by \citet{Wahhaj2014} to $x_c=1.2\pm0.1$AU is also detected in our data with a larger uncertainty, $x_c=0.8^{+3.5}_{-3.0}$AU. This uncertainty, given at $3\sigma$, includes the error on the star centre but this contribution is not dominant because these observations are done without a coronagraph in contrast to the HST/STIS observations. 

\begin{table*}
%\caption{Measurements of the morphology of the disc for the \Lp{} intensity image, in four different scenarii: an anisotropic centered disc (first row), an anisotropic disc centered along its minor-axis (second row), an isotropic off-centered disc (third row) and an anisotropic off-centered disc (fourth row). The uncertainty is given at $1\sigma$.}             % title of Table
\caption{Measurements of the morphology of the disc from the Ks polarised intensity image and Lp intensity image. The parameter $x_c$ is the offset along the major-axis (negative means the SW side is closer to the star), and $y_c$ is the deprojected offset along the minor-axis (negative means the NW side is closer to the star). The uncertainty is given at $3\sigma$ and includes the uncertainty from the measurement and the star position. In \Lp{} we investigated two scenarios: an isotropic disc ($g$ set to 0) and an anisotropic disc (g as a free parameter).}             % title of Table
\label{tab_disc_param_Lp}      % is used to refer this table in the text
\centering                          % used for centering table
\begin{tabular}{c c | c c c c c c | c}        % centered columns (4 columns)
\hline\hline                 % inserts double horizontal lines
Filter & polarisation & PA & $r_0$ (AU) & $i$ & $x_c$(AU) & $y_c$ (AU) & $g$ & $\chi^2$ \\    % table heading 
\hline                        % inserts single horizontal line
Ks & Yes & ${26.7^{\circ}}^{\pm1.6^\circ}$ & $75.3_{-2.2}^{+2.0}$ & ${75.5^{\circ}}^{+1.3^\circ}_{-1.7^\circ}$ & $0.8^{+3.5}_{-3.0}$ & $-5.8^{\pm8.3}$ & NA & 1.6 \\     
\Lp{} & No &  ${26.9^{\circ}}^{\pm1.5^\circ}$ & $74.9^{\pm2.1}$ & ${76.0^{\circ}}^{\pm1.5^\circ}$ & $1.9^{\pm3.0}$ & $-3.5^{\pm9.6}$ & 0\tablefootmark{a} & 0.93 \\      % 
\Lp{} & No & ${26.9^{\circ}}^{+1.5^\circ}_{-1.2^\circ}$ & $74.8_{-1.8}^{+2.1}$ & ${75.8^{\circ}}_{-1.8^\circ}^{+1.5^\circ}$ & $2.0^{\pm3.0}$ & $-3.5^{+9.3}_{-9.9}$ & $0.01^{+0.15}_{-0.18}$ & 0.93 \\                                  %inserts single line
\end{tabular}
\tablefoot{
\tablefoottext{a}{Fixed parameter.}
}
\end{table*}
 
\subsubsection{Constraints from unpolarised observations at \Lp{}}
\label{sec_Lp_constraints}

To measure the morphology of the disc in the \Lp{} band, we keep the same parametric model as introduced previously and model now the anisotropy of scattering with a Henyey-Greenstein phase function instead of an ad hoc smooth function. In that case, the Henyey-Greenstein g coefficient is an additional free parameter that can be used to quantify the anisotropy of scattering. Because the star subtraction is performed using ADI, the disc can be significantly self-subtracted \citep{Milli2012}, and we implement the forward-modelling technique already described in \citet{Milli2014} for $\beta$ Pictoris  to retrieve the best disc parameters. The results are shown in Table \ref{tab_disc_param_Lp}.  Our results agree with the values already published in \citet{Lagrange2012_HR4796}. They are compatible with a disc scattering light isotropically.  The offset $y_c=-3.5\pm9.6$AU  detected along the semi-minor axis explains the NW/SE asymmetry without the need for anisotropic scattering.  An offset of $y_c=-1.15\pm0.16$AU along the semi-minor axis was already detected by \citet{Thalmann2011} in the H band. 

%We derive a total \Lp{} magnitude of the disc of 12.4.

\subsection{Photometry in the Ks band}

\subsubsection{Polarised light}

In polarised light, we find a polarised flux of $13.2 \pm0.1 \text{mag/arcsec}^2$ and $12.3 \pm0.1 \text{mag/arcsec}^2$ for the NE and SW ansae of the disc, respectively, at a projected separation of $1.05\arcsec$ or $76.7AU$.  The uncertainty is computed as three times the azimuthal root mean square in the radial Stokes and does not take systematic errors that could could arise from a bad calibrations of the instrumental polarisation into account . The polarised flux is maximal along the semi-minor axis, on the north-west (NW) side. 

\subsubsection{Unpolarised light}

 Below $0.6\arcsec$, the noise dominates the image, so we restrict the photometry of the disc to an elliptical aperture beyond $0.6\arcsec$ as done by \citet{Debes2008}. We use an isotropic model of the disc to correct for the flux loss due to PCA and for the disc flux not measured in the aperture. We find a total disc flux of $3.7\pm0.26$ mJy compatible with the $3.5\pm0.2$ mJy measured by \citet{Debes2008}. 

\subsection{Polarised fraction}

\begin{figure}
   \centering
   \includegraphics[width=\hsize]{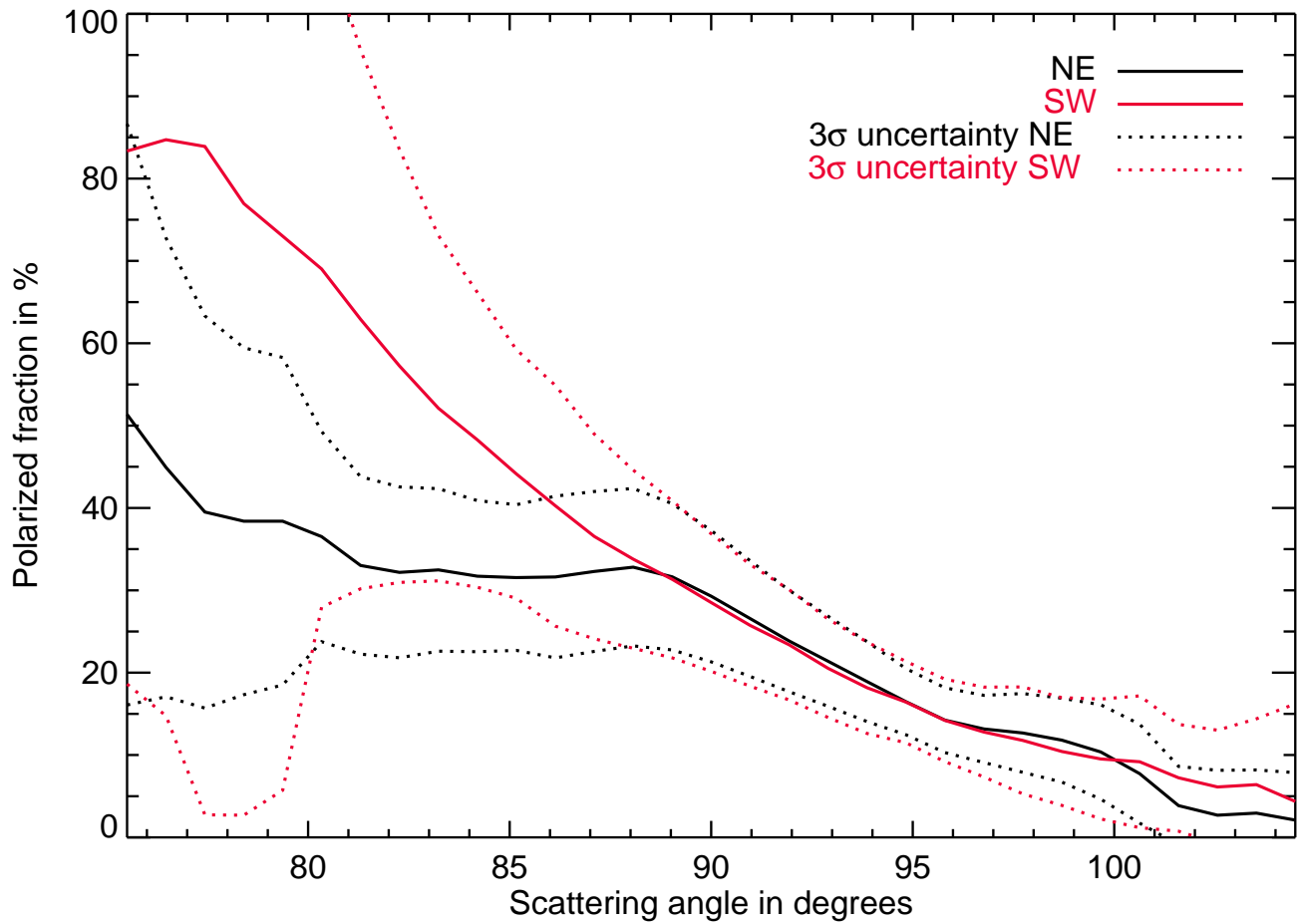}
      \caption{Polarised fraction as a function of the scattering phase angle, measured from NaCo Ks polarised intensity and NICMOS intensity. A phase angle of $0^\circ$ indicates the NW side and  $90^\circ$ corresponds either to the NE or SW ansa.}
         \label{fig_p}
   \end{figure}

The polarised fraction (Fig. \ref{fig_p}) is computed as the ratio of the NaCo pI image over the NICMOS I image. It can only be computed reliably within $15^\circ$ of the ansae because of the residual noise affecting the NICMOS image beyond this range. The polarised fraction reaches $29.3\%\pm8.0\%$ and $28.5\%\pm8.4\%$ in the NE and SW ansae, and is increasing continuously westwards from both ansae for at least $15^\circ$. If we assume that the side inclined towards us is the NW side (see discussion in section \ref{sec_discussion}), then it means that the peak polarisation fraction occurs below a scattering angle of $90^\circ$ and  we can state with a $99.7\%$
confidence level that the polarised fraction is above 22\% for scattering angles between $80^\circ$ and $90^\circ$ . Although the polarised fraction is very similar on both sides of the disc above a scattering angle of $90^\circ$, stronger values are measured below $88^\circ$ for the SW side. However our large error bars are still compatible with a similar behaviour for dust on both sides of the disc.

\subsection{Disc colour}
\label{sec_colour}

From our measurements at two different wavelengths, we calculate a colour index defined as 
\begin{equation}
\frac{I_{disc,L^\prime}/I_{\star,L^\prime}}{I_{disc,{Ks}}/I_{\star,{Ks}}}
\label{eq_color}
\end{equation}
where $I_{\star}$ is the star flux density at the specified wavelength and $I_{disc}$ is the disc flux density measured in the ansae for the reason explained previously. 
We find a value of $1.1 \pm 0.6$. Although the error bar is relatively large, this extends the conclusion by \citet{Debes2008} of a dust spectral reflectance flattening beyond \unit{1.6}{\micro\meter}. Their conclusion was  valid only up to \unit{2.2}{\micro\meter}.  We provide  a new photometric point at \Lp{} and measure that the flattening is valid up to \unit{3.8}{\micro\meter}.

\section{Modelling}
\label{sec_modelling}

In the previous section, we reveal a very puzzling brightness inversion between the NW and SE sides of the disc and present new observational constraints. To test these features against our current understanding of light scattering by dust particles, we model the radiative transfer in the disc and generate synthetic scattered light images.

Contrary to what was assumed before \citep{Schneider2009}, our baseline scenario to explain these features, which is discussed in section \ref{sec_discussion}, is the following: the NW side of the ring is inclined towards the Earth. This scenario is called $H_1$ and the alternative hypothesis is called $H_2$.

\subsection{Approach, assumptions, and parameter space exploration}

Within the scope of this paper, we investigate whether a disc model can simultaneously fit the SED and explain the scattered light images. Extensive modelling work of the SED of the disc around HR 4796A has already been performed in past studies (e.g. \citealt{Augereau1999,Li2003}) to constrain the dust geometrical extension and properties. Our new observations can constrain the cold outer dust component of their model, which is responsible for the far-infrared excess beyond \unit{25}{\micro\meter}. The best model able to reproduce the far-infrared emission (model \#13 of \citealt{Augereau1999}) assumes a population of dust grain with sizes from $a_{min}=\unit{10}{\micro\meter}$ to $a_{max}=\unit{1}{\meter}$ with a $s=-3.5$ power-law exponent \citep{Dohnanyi1969}, and composed of porous silicates coated by an organic refractory mantle with water ice partially filling the holes due to porosity. We keep the same notation as in their paper, namely a total porosity $P_{wH_2O}$, a fraction of vacuum removed by the ice $p_{H_2O}$, a silicates over organic refractory volume fraction $q_{Sior}$. The porosity of the grain once the ice has been removed is written $P$. Model \#13 of  \citet{Augereau1999} used a porosity $P$ of 59.8\% and $p_{H_2O}=3\%$.
We reproduced their model with the radiative transfer code MCFOST \citep{Pinte2006,Pinte2009}, we used the effective medium theory (EMT) to derive the optical indices of the grains, and the Mie theory \citep{Mie1908} to compute their absorption and scattering properties. The EMT-Mie theory assumes porous spheres, which is probably not realistic for fluffy aggregates larger than \unit{1}{\micro\metre} and can lead to unrealistic scattering properties \citep{Voshchinnikov2007}. Therefore, we  also investigated a statistical approach, known as the distribution of hollow spheres \citep[DHS,][]{Min2005}, with an irregularity parameter $f_\text{max}=0.8$. This approach averages the optical properties of hollow spheres over the fraction of the central vacuum. It was proven successful to reproduce the scattered light behaviour of randomly oriented irregular quartz particles. In both cases, Mie and DHS, we use the full-scattering matrix to compute the synthetic intensity and polarisation maps. This is a major difference from \citet{Augereau1999}, who purposedly adopted the empirical Henyey-Greenstein phase function because an incompatibility between the SED and the scattered light images predicted by the Mie theory was already seen at the time. The Mie phase function for large grains predicted by the SED modelling indeed exhibits  a very pronounced peak for small scattering angles.
We assume a population of dust grains, located in an annulus centred about the star and inclined by $75.8^\circ$, whose radial volume density in the midplane follows a piecewise power-law of exponent 9.25 before $R_c=74.2AU$ and $-12.5$ after this radius. For the vertical distribution, we assume a Gaussian profile of scale height $H=1AU$ at $R_c$. This represents a slight revision with respect to \citet{Augereau1999}, who used a wider vertical profile less compatible with the new observations. We use the aspect ratio $H/R_c$  0.013, similar to Fomalhaut and slightly lower than the "natural" aspect ratio of $0.04\pm0.2$ derived by \citet{Thebault2009} for debris discs. We initially generated the models  with a small dust mass to keep the synthetic discs optically thin, and then scaled up to match the measured SED above \unit{20}{\micro\meter}. This approach is correct under two conditions: 1)  if the disc is optically thin along the line of sight, 2) if the SED above \unit{20}{\micro\meter} is dominated by the contribution of the cold annulus seen in scattered light. We validated the condition 1) by re-running the simulation for the best models with the corrected dust mass (see section \ref{sec_optical_thickness}). Conditon 2) is consistent with the findings of \citet{Augereau1999} who showed that the cold component contributes to 90\% of the measured excess at \unit{20.8}{\micro\meter}.

There are five remaining free parameters of the models, including $(a_{min},P,p_{H_2O},q_{Sior},s)$. We adopt a coarse sampling resulting in a grid of 15600 models that encompass previous estimates proposed by \citet{Augereau1999} to explain the SED, early resolved scattered light and thermal observations. The details of this grid are summarized in Table \ref{tab_grid}. This grid represents the most extensive modelling ever realised for this object.

\begin{table}
\caption{Grid of parameters for the 15600 models generated.}             % title of Table
\label{tab_grid}      % is used to refer this table in the text
\centering                          % used for centering table
\begin{tabular}{p{1.7cm} p{1cm} p{1cm} p{1cm} p{1.3cm}}        % centered columns (4 columns)
\hline\hline                 % inserts double horizontal lines
Parameter & Min. value & Max. value & $N_\text{sample}$ & Sampling  \\    % table heading 
\hline    
Scattering theory &  \multicolumn{2}{c}{Mie / DHS} & / & / \\   % inserts single horizontal line
 $a_{min}$ ({\textmu}m) & 0.1 & 100 & 13 & log. \\      % inserting body of the table
$p_{H_2O}$ (\%) & 1 & 90 & 5 & log. \\      % inserting body of the table
$P$ (\%) & 0 & 80 & 5 & linear \\      % inserting body of the table
$q_{Sior}$ & 0 & 1 & 6 & linear \\
$s$ & -2.5 & -5.5 & 4 & linear \\
\hline                                   %inserts single line
\end{tabular}
\end{table}

\begin{table}
\caption{Infrared and submillimeter flux density used in the SED fitting procedure. These measurement are posterior to those used in the  early modelling by \citet{Augereau1999}}           % title of Table
\label{tab_SED}      % is used to refer this table in the text
\centering                          % used for centering table
%\begin{tabular}{c c c c c}        % centered columns (4 columns)
\begin{tabular}{ p{0.6cm} p{1cm} p{1cm} p{1.6cm} p{1.7cm}}        % centered columns (4 columns)
\hline\hline                 % inserts double horizontal lines
$\lambda$ ({\textmu}m) & Flux density (Jy)  & Error (Jy) & Instrument & Source \\
\hline                        % inserts single horizontal line
20.8 &1.813 &0.17 &MIRLIN & \citet{Koerner1998} \\      
24 &3.03 &0.303 &SPITZER MIPS & \citet{Low2005} \\  
65 &6.071 &0.313 &AKARI   & \citet{AkariCat2010}\\ %
70 &4.98 &0.131 &HERSCHEL PACS  &\citet{Riviere-Marichalar2013} \\
90 &4.501 &0.186 &AKARI   & \citet{AkariCat2010}\\ %\citet{Liu2013} \\
100 &3.553 &0.097 &HERSCHEL PACS &\citet{Riviere-Marichalar2013} \\ 
160 &1.653 &0.068 &HERSCHEL PACS &\citet{Riviere-Marichalar2013} \\ 
870 &0.0215 &0.0066 &LABOCA APEX & \citet{Nilsson2010} \\
\hline                                   %inserts single line
\end{tabular}
\end{table}

\subsection{Goodness of fit estimators}
\label{sec_estimators}

Firstly, to find the best models that can explain the SED, we choose as an estimator for the goodness of fit, a reduced Chi square $\chi^2_\text{SED}$ between the synthetic SED, and eight SED measurements above \unit{20}{\micro\meter} listed in Table \ref{tab_SED}. Secondly, to find the best models that can explain the scattered light observations, we build several additional estimators for the goodness of fit. These include:
\begin{enumerate}
\item two reduced Chi square $\chi^2_{p,H_i}$ ($i \in \{1,2\}$) of the polarised fraction $p$, computed from 31 measurements corresponding to scattering angles between $75^\circ$ and $105^\circ$. We average the measurements between the NE and SW ansae. Because we want to test the hypothesis $H_1$ (NW inclined towards the Earth) and $H_2$ (SE inclined towards the Earth), there is a different Chi square $\chi^2_{p,H_1}$ and $\chi^2_{p,H_2}$ for each scenario.
\item two reduced Chi square $\chi^2_{p\times\phi,H_i}$ ($i \in \{1,2\}$) of the polarised phase function $p\times \phi$. They are computed using all measurements where the disc is detected in polarised light namely from $14.8^\circ$ to $105^\circ$ for the scenario $H_1$ and from $165.2^\circ$ down to $75^\circ$ for the scenario $H_2$. They also include upper limits set by the observations between $105^\circ$ and $165.2^\circ$ for $H_1$ (between $14.8^\circ$ and $75^\circ$ for $H_2$, respectively).
\item a reduced Chi square $\chi^2_\text{albedo}$ of the effective albedo measured in the ansae.
\item a reduced Chi square $\chi^2_\text{colour}$ of the colour index defined in Eq. \ref{eq_color}.
\end{enumerate}
These four observables are independent, we can therefore combine them into a single reduced Chi square that we use as an estimator for the goodness of fit of all our scattered light observables, defined as 
\begin{equation}
\label{eq_chi2_scat_light}
\chi^2_{\text{scat. light},H_i} = \chi^2_{p,H_i} +  \chi^2_{p\times\phi,H_i} + \chi^2_\text{colour} + \chi^2_\text{albedo}
.\end{equation}

The phase function $\phi$ is not part of these four observables because it is not independent from them and can be deduced from the polarised phase function and the polarised fraction.  For the purpose of the analysis, however, it is still very instructive to study the agreement of the models with the phase function alone. This is why we introduce additionally two reduced Chi square $\chi^2_{\phi,H_i}$  ($i \in \{1,2\}$) of the phase function $\phi$, computed from 31 measurements of $\phi$ at scattering angles between $75^\circ$ and $105^\circ$ for the scenarios $H_1$ and $H_2$.

Lastly, we build an overall Chi square combining the constraints from the SED and the scattered light to do a global fitting of all observables simultaneously, which yields
\begin{equation}
\chi^2_{\text{overall},H_i} = \chi^2_{\text{scat. light},H_i}  + \chi^2_\text{SED}
.\end{equation}

\begin{figure*}
   \centering
   \includegraphics[width=\hsize]{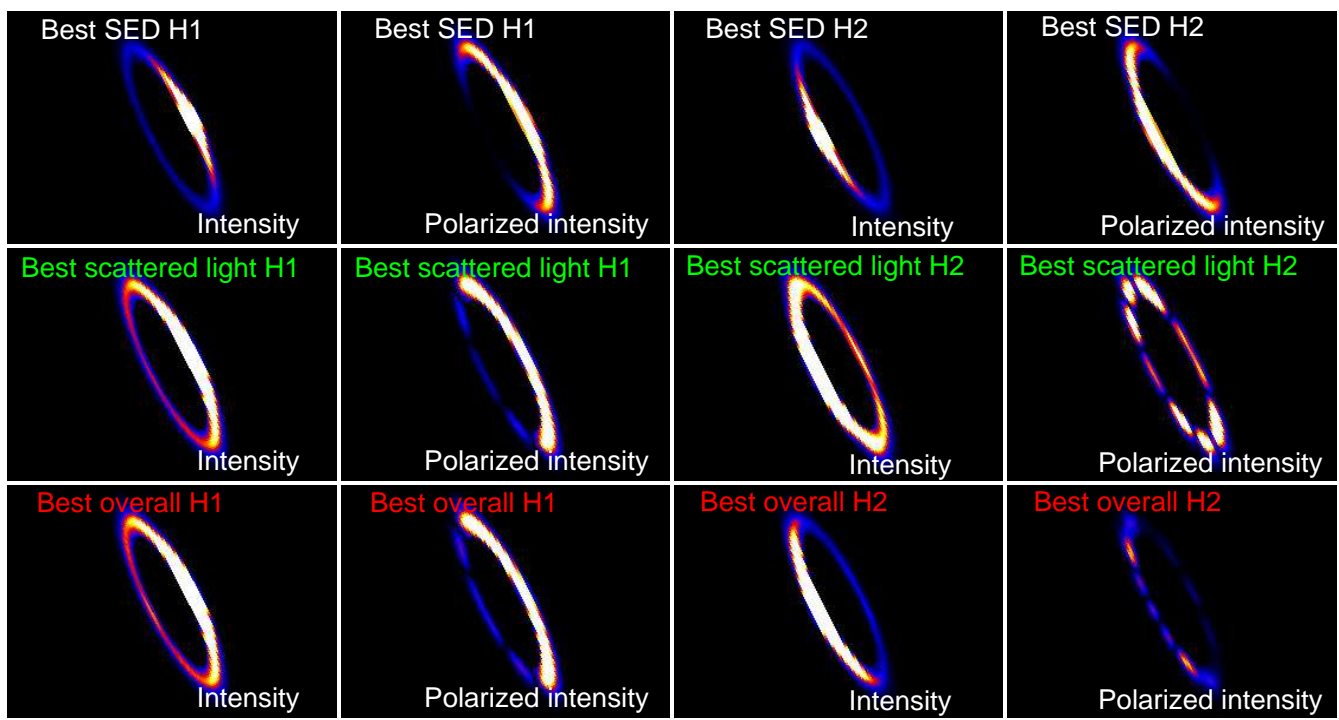}
      \caption{Unpolarised and polarised intensity of the best SED models (first row), the best scattered light models (second row), and the best overall models (last row), detailed in Table \ref{tab_chi2_Mie_DHS}. The  images on the left half refer to best models in the scenario $H_1$ and the  images on the right half refer to the best models in the scenario $H_2$. The colour scale is linear for all images. The colour range is the same for all intensity images and is four times smaller for the polarised images.}
       \label{fig_prediction_I_pI_best_models}
   \end{figure*}

\begin{figure*}
   \centering
   \includegraphics[width=\hsize]{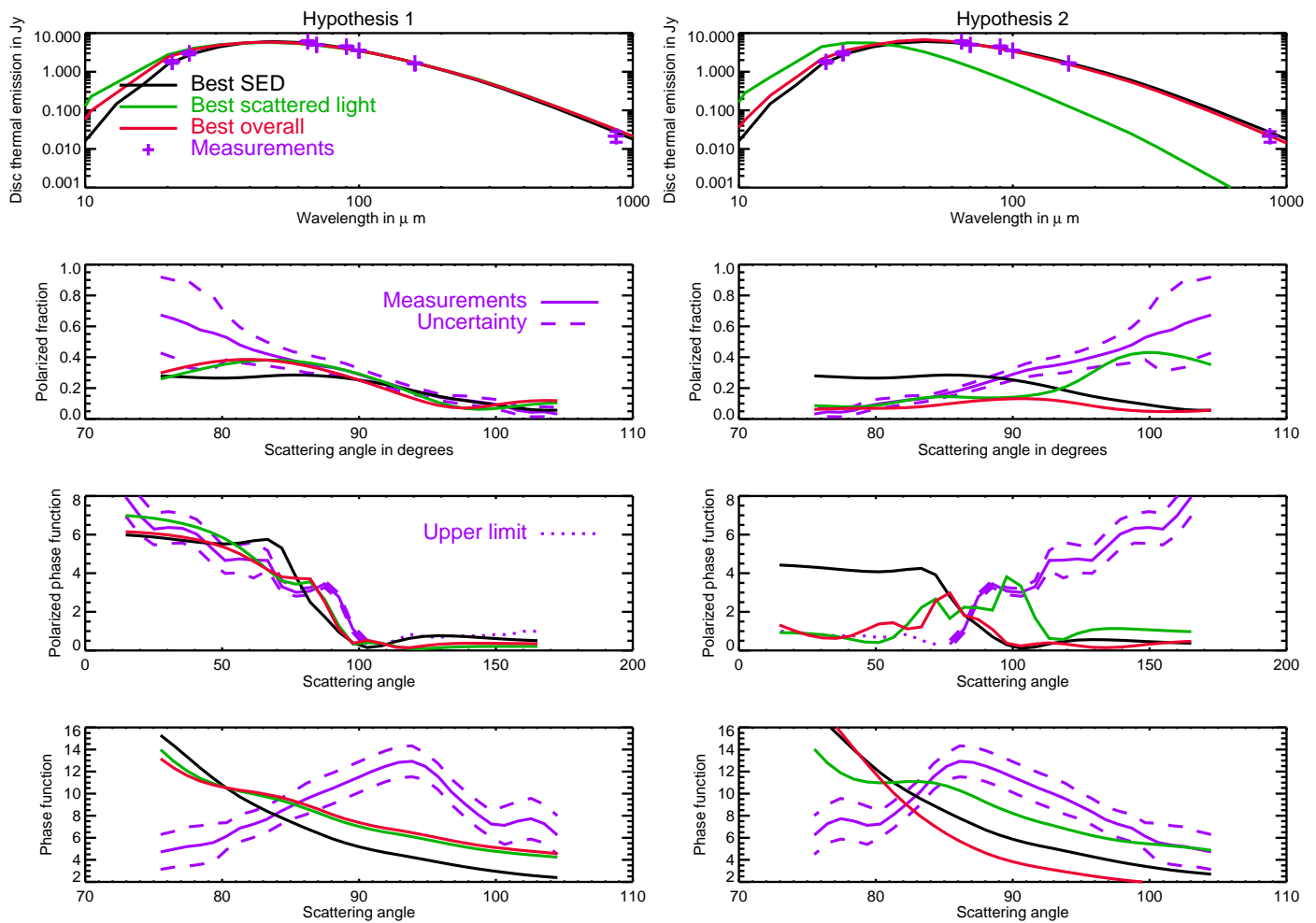}
      \caption{Comparison between the measurements (in purple) and the predictions of the best model for the SED (in black), the scattered light (in green), or overall (in red) with respect to four observables: the SED (first row), the polarised fraction (second row), the polarised phase function (third row), and the phase function (last row). The left column shows the best models in the scenarios $H_1$ (NW side towards the Earth) and the right column shows the best models in the scenario $H_2$. }
       \label{fig_theory_vs_measurements}
   \end{figure*}

The Chi square values that maximise the goodness of fit estimators  are detailed in Table \ref{tab_chi2_Mie_DHS}, along with the parameters associated with these models. An illustration of the corresponding scattered light synthetic images is given in Fig. \ref{fig_prediction_I_pI_best_models} and a quantitative comparison between the measurements and the predictions is shown in Fig. \ref{fig_theory_vs_measurements}. These results are discussed in section  \ref{sec_modelling_results_sed} and  \ref{sec_modelling_results_scat_light}.

\begin{sidewaystable*}
\caption{Goodness of fit estimates and corresponding parameters for the best models with respect to the SED or the scattered light observables.}            
\centering          
\label{tab_chi2_Mie_DHS}      
%\begin{tabular}{ c | C{1.05cm} C{1.05cm} C{1.05cm} C{1.05cm} C{1.05cm} C{1.05cm} C{1.05cm} C{1.05cm} C{1.1cm} C{1.1cm} C{1.1cm} C{1.1cm} C{1.1cm} }     % 7 columns 
\begin{tabular}{ c | c c c c c c c c c c c c c }     % 7 columns 
\hline\hline       
%  & best SED & best p\tablefootmark{a} ($H_1$) & best p ($H_2$) & best $p\times\phi$\tablefootmark{b} ($H_1$) & best $p\times\phi$ ($H_2$) & best $\phi$\tablefootmark{c} ($H_1$) & best $\phi$ ($H_2$) & best colour & best albedo & best scat. light\tablefootmark{d} ($H_1$) & best scat. light ($H_2$) & best overall\tablefootmark{e} ($H_1$) & best overall ($H_2$) \\
  & best   & best   & best   & best   & best   & best   & best   & best   & best   & best   & best   & best   & best \\
  & SED & p\tablefootmark{a} ($H_1$) & p ($H_2$) & $p\times\phi$\tablefootmark{b} & $p\times\phi$ & $\phi$\tablefootmark{c} & $\phi$ & colour & albedo & scat. & scat. & overall\tablefootmark{e} & overall \\
  &  &  &  & ($H_1$) & ($H_2$) &  ($H_1$) & ($H_2$) &  & &  light\tablefootmark{d} & light  &  ($H_1$) &  ($H_2$) \\
  &  &  &  &  & &  & &  &  & ($H_1$) & ($H_2$) &   & \\
  \hline                  
Theory & DHS & Mie & Mie & DHS & Mie & DHS & DHS & Mie & DHS & Mie & Mie & Mie & Mie \\
s & -3.5 & -4.5 & -4.5 & -5.5 & -5.5 & -5.5 & -5.5 & -4.5 & -2.5 & -3.5 & -5.5 & -3.5 & -3.5 \\
$q_\text{Sior}$ & 0.2 & 1.0 & 0.2 & 1.0 & 1.0 & 0.0 & 0.4 & 1.0 & 0.0 & 0.2 & 0.8 & 1.0 & 0.6 \\
$p_{H_2O}$ & 3.1\% & 9.5\% & 9.5\% & 90.0\% & 90.0\% & 1.0\% & 90.0\% & 1.0\% & 29.2\% & 1.0\% & 90.0\% & 1.0\% & 9.5\% \\
$a_\text{min}$ & 1.78 & 1.78 & 1.78 & 10.00 & 0.56 & 10.00 & 10.00 & 1.00 & 10.00 & 1.00 & 1.00 & 1.00 & 1.78 \\
P & 20.0\% & 20.0\% & 80.0\% & 40.0\% & 0.1\% & 0.1\% & 60.0\% & 60.0\% & 0.1\% & 20.0\% & 0.1\% & 0.1\% & 60.0\% \\
\hline                  
$\chi^2_\text{SED}$ & \textbf{1.7} & 158.6 & 359.8 & 63.3 & 126.5 & 32.9 & 90.5 & 394.6 & 77.3 & 11.6 & 94.4 & 6.6 & 5.4 \\
$\chi^2_{p,H_1}$ & 66.8 & \textbf{0.4} & 82.4 & 111.4 & 114.3 & 52.1 & 132.0 & 24.5 & 54.5 & 1.3 & 21.1 & 2.2 & 10.2 \\
$\chi^2_{p,H_2}$ & 249.1 & 75.3 & \textbf{0.1} & 408.5 & 25.8 & 361.5 & 411.1 & 7.7 & 364.5 & 38.2 & 2.5 & 41.4 & 10.4 \\
$\chi^2_{p\times\phi,H_1}$ & 32.9 & 22.3 & 15.2 & \textbf{2.1} & 22.1 & 11.8 & 2.4 & 25.5 & 8.0 & 7.4 & 23.5 & 9.6 & 17.4 \\
$\chi^2_{p\times\phi,H_2}$ & 249.1 & 75.3 & 0.1 & 408.5 & \textbf{0.8} & 361.5 & 411.1 & 7.7 & 364.5 & 38.2 & 3.7 & 41.4 & 10.1 \\
$\chi^2_{\phi,H_1}$ & 10.0 & 12.3 & 30.8 & 2.8 & 28.7 & \textbf{1.7} & 4.0 & 30.4 & 4.0 & 11.0 & 25.8 & 9.7 & 26.8 \\
$\chi^2_{\phi,H_2}$ & 5.6 & 6.5 & 24.1 & 2.4 & 21.8 & 3.5 & \textbf{1.8} & 23.9 & 2.1 & 5.5 & 18.5 & 4.6 & 19.7 \\
$\chi^2_\text{colour}$ & <0.1 & 5.2 & <0.1 & 1.2 & 0.1 & 1.4 & 1.0 & \textbf{<0.1} & 0.1 & 1.4 & <0.1 & 1.4 & 0.1 \\
$\chi^2_\text{albedo}$ & 24.9 & 4.5 & 9.5 & 37.0 & 3.9 & 36.6 & 37.0 & <0.1 & \textbf{<0.1} & 4.6 & 4.4 & 4.0 & 10.4 \\
$\chi^2_{\text{scat. light},H_1}$ & 124.6 & 32.4 & 107.2 & 151.7 & 140.3 & 101.9 & 172.4 & 50.0 & 62.6 & \textbf{14.6} & 49.0 & 17.1 & 38.2 \\
$\chi^2_{\text{scat. light},H_2}$ & 2061.7 & 381.9 & 309.4 & 873.9 & 112.7 & 1314.6 & 827.9 & 457.5 & 564.2 & 364.1 & \textbf{55.8} & 306.1 & 117.5 \\
$\chi^2_{\text{overall},H_1}$ & 125.3 & 181.2 & 457.4 & 176.7 & 262.8 & 96.8 & 224.8 & 444.6 & 139.8 & 20.2 & 139.0 & \textbf{23.7} & 43.6 \\
$\chi^2_{\text{overall},H_2}$ & 2063.4 & 540.5 & 669.2 & 937.2 & 647.6 & 1347.4 & 918.3 & 852.1 & 641.6 & 375.7 & 523.1 & 312.7 & \textbf{122.8 }\\
\hline
\end{tabular}
\tablefoot{
\tablefoottext{a}{Best model of the polarised fraction.}
\tablefoottext{b}{Best model of the polarised phase function.}
\tablefoottext{c}{Best model of the phase function.}
\tablefoottext{d}{Best model of all independent scattered light observables: polarised fraction, polarised phase function, effective albedo and colour index.}
\tablefoottext{e}{Best model of the SED and the scattered light.}
}
\end{sidewaystable*}

\subsection{Bayesian formalism}

To provide an estimate of the range of acceptable models for each goodness of fit estimators, we carry out a Bayesian analysis \citep[e.g.][]{Pinte2008,Duchene2010}.

Each model is assigned a probability that the data are drawn from the model parameters. In our case, we do not have any a priori information on these parameters, we therefore assumed a uniform prior, corresponding to a uniform sampling of our parameters by our grid. We used for most parameters a linear sampling of the free parameters of our model (see Table \ref{tab_grid}), except for the minimum grain size and the fraction of vacuum occupied by the ice, for which a logarithmic sampling was more natural. Under this uniform prior assumption, the probability $\Psi$ that  the data corresponds to a given parameter set is given by
\begin{equation}
\Psi=\Psi_0\text{exp}\left( -\frac{\chi^2}{2}\right)
,\end{equation}
$\Psi_0$ is a normalisation constant introduced so that the sum of probabilities over all models is unity. The probability given here is only valid within the framework of our modelling and parameter space.

\begin{figure*}
\centering
\includegraphics[width=\hsize]{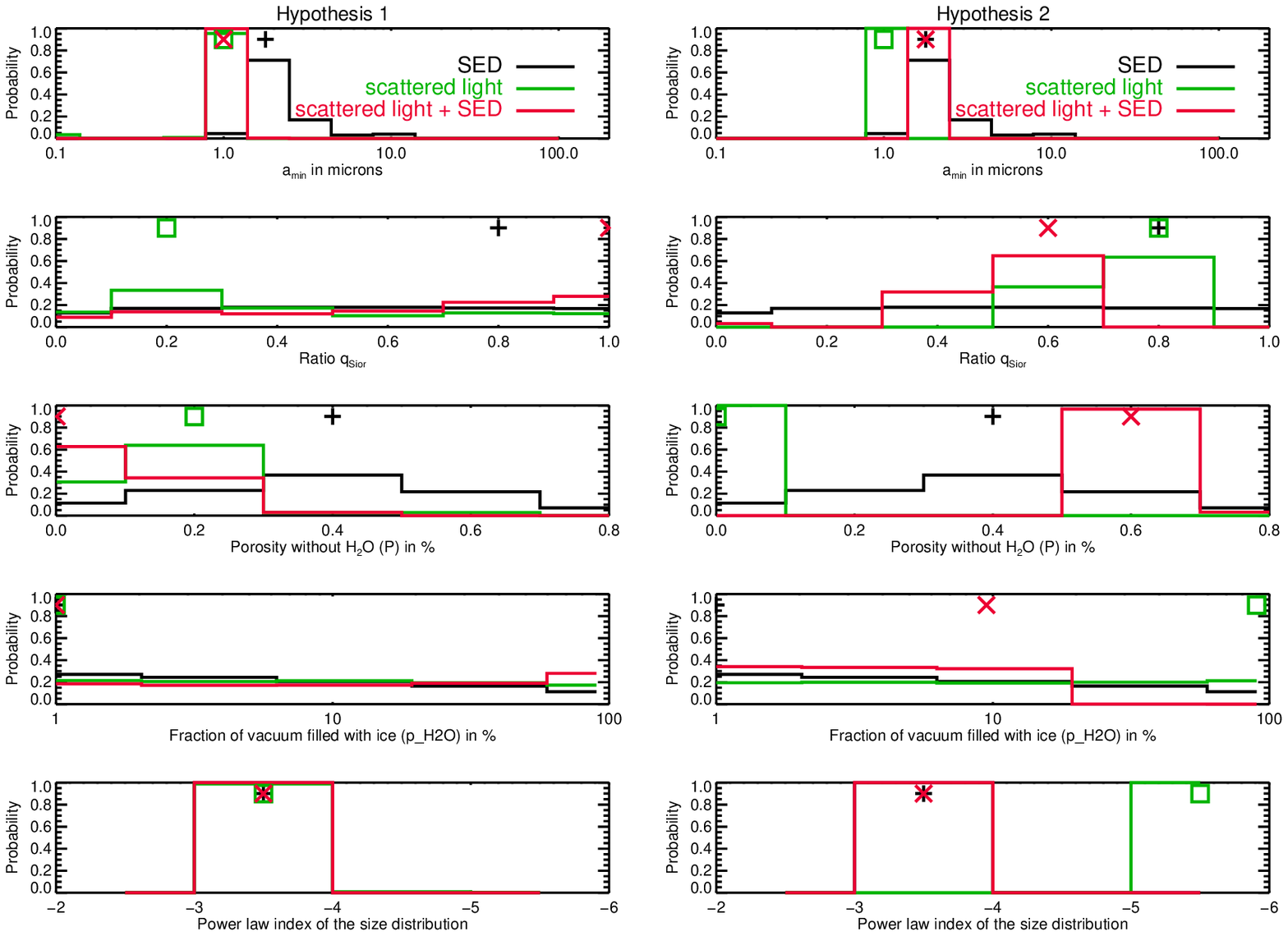}
\caption{Marginal distributions of the five free parameters of our model, based on fitting the SED (black histogram), the scattered light images (polarised fraction, polarised phase function, colour index and effective albedo, green histogram), or both (red histogram). The symbols (crosses and squares) indicate the value of the best SED, scattered light or overall model. These distributions are derived from models created using the Mie theory.}
\label{fig_marginal_PDF_Mie}
\end{figure*}

Fig. \ref{fig_marginal_PDF_Mie} shows the inferred probability distributions for each of our five free parameters, after marginalisation against all four parameters. It is shown here using the Mie theory because the best scattered light models and overall models are obtained with this theory, but in Appendix \ref{App_marginal_PDF_DHS} we provide the probability distributions obtained using the DHS theory.

\subsection{Constraints brought by the SED fitting alone}
\label{sec_modelling_results_sed}

The best SED model has a $\chi^2_\text{SED} $ of 1.7, as shown in the first column of Table \ref{tab_chi2_Mie_DHS}. This model points towards grains with a minimum size of \unit{1.8}{\micro\meter}. This value is somehow smaller than the \unit{10}{\micro\meter} proposed by \citet{Augereau2001}. We find an improvement using a slightly higher carbon content and smaller porosity than the values $q_{Sior}=0.47$ and $P=59.8\%$  set by \citet{Augereau1999}. We attribute these differences to our updated SED measurements. The fit of the SED is shown in the first two panels in Fig. \ref{fig_theory_vs_measurements} and the corresponding intensity and polarised images are shown in the first row of images in Fig. \ref{fig_prediction_I_pI_best_models}.

\subsection{An incompatibility between the scattered light observables}
\label{sec_modelling_results_scat_light}
%\label{sec_modelling_results}

Among our grid, we  find models that provide a good fit to each scattered light observables taken individually, with Chi square values below 2.1. However, when we combine all four independent scattered light observables together, the best model only yields a Chi square $\chi^2_{\text{scat. light},H_i} $ of 14.6 and 55.8 in the scenario $H_1$ and $H_2$ , respectively. To illustrate  this mismatch visually, we show the synthetic scattered light images corresponding to those best models in Fig. \ref{fig_prediction_I_pI_best_models} (second row) and the measurements and model predictions for four observables are plotted in Fig. \ref{fig_theory_vs_measurements} (green lines). The brightness inversion between polarised and unpolarised light cannot be reproduced by any of these models. As an example, Fig. \ref{fig_theory_vs_measurements} shows that under the assumption $H_1$, the best scattered light model reproduces  the polarisation measurements well, but the phase function is incompatible with the measurements. Under the assumption $H_2$, the measured and predicted phase function agree better but the polarised phase function cannot be explained. Therefore, the scattered light best models represent a trade-off that fail to reproduce  all the independent scattered light observables simultaneously. Furthermore, taking the SED into account only makes matters worse. The overall best models (last two columns of Table \ref{tab_chi2_Mie_DHS}) have indeed a $\chi^2_{\text{scat. light},H_i}$ of 17.1 and 117.5 respectively, while the $\chi^2_\text{SED}$ is more than three times above the best SED model.

We will now focus on this incompatibility and analyse specifically two different features at the heart of this problem: first, the change in the brightest side between polarised and unpolarised light, and second, the decrease in the anistropy of scattering with increasing wavelength. Based on our Bayesian analysis, we will show that there are no solutions that can fully explain these features in our  exhaustive parameter space.

\subsubsection{Change in the brightest side between polarised and unpolarised light}
\label{sec_brightness_inversion}

\begin{figure}
   \centering
   \includegraphics[width=\hsize]{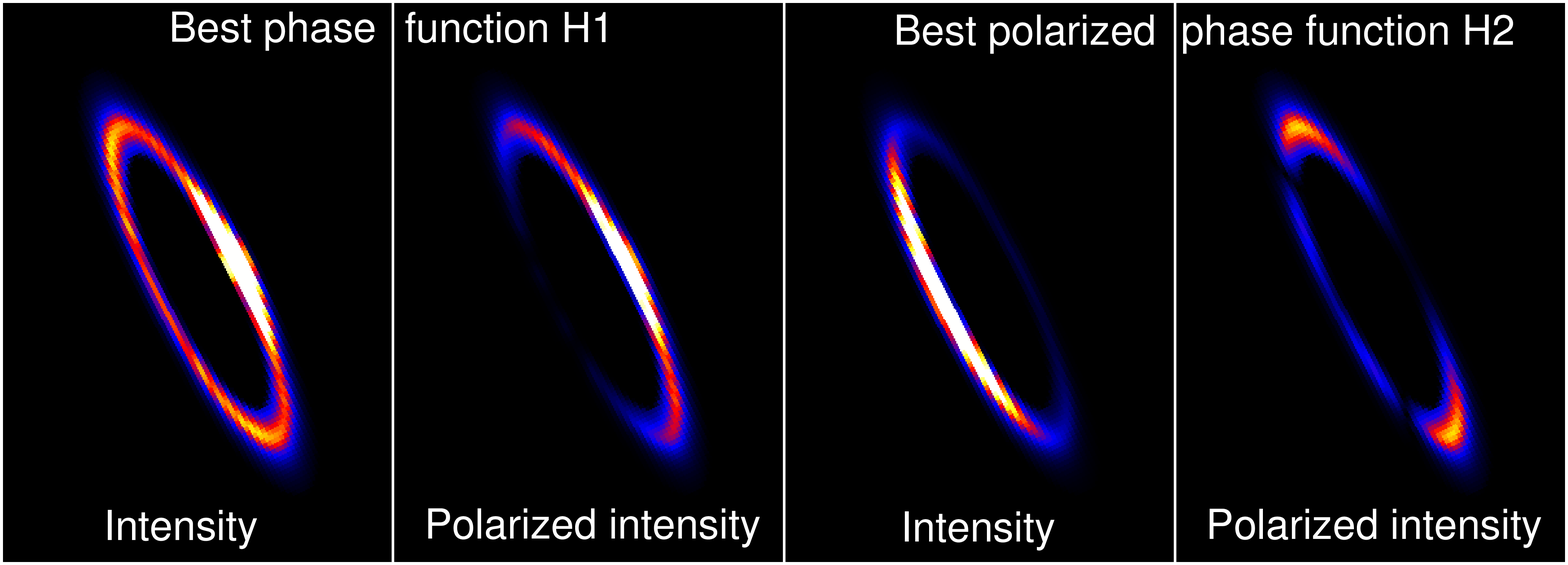}
      \caption{Ks scattered light images of the best model for the phase function in the scenario $H_1$ (left images) and the best model for the polarised phase function in the scenario $H_2$ (right images). The colour scale is linear and identical for the two images.}
       \label{fig_scattered_light_best_phase_model}
   \end{figure}

Among the best SED, scattered light, and overall models presented in Fig. \ref{fig_prediction_I_pI_best_models}, we cannot find any model where the brightest side changes between polarised and unpolarised light. Indeed, those models fail to fit  the phase function and the polarised phase function simultaneously, as visible in Fig. \ref{fig_theory_vs_measurements}. 
Because the forward side is always brighter in intensity in all our models, the models that can best reproduce this inversion in brightness within our grid are either the best models for the phase function in  scenario $H_1$, or the best models for the polarised phase function in  scenario $H_2$. They are presented in Fig. \ref{fig_scattered_light_best_phase_model}. 

\begin{figure}
   \centering
   \includegraphics[width=\hsize]{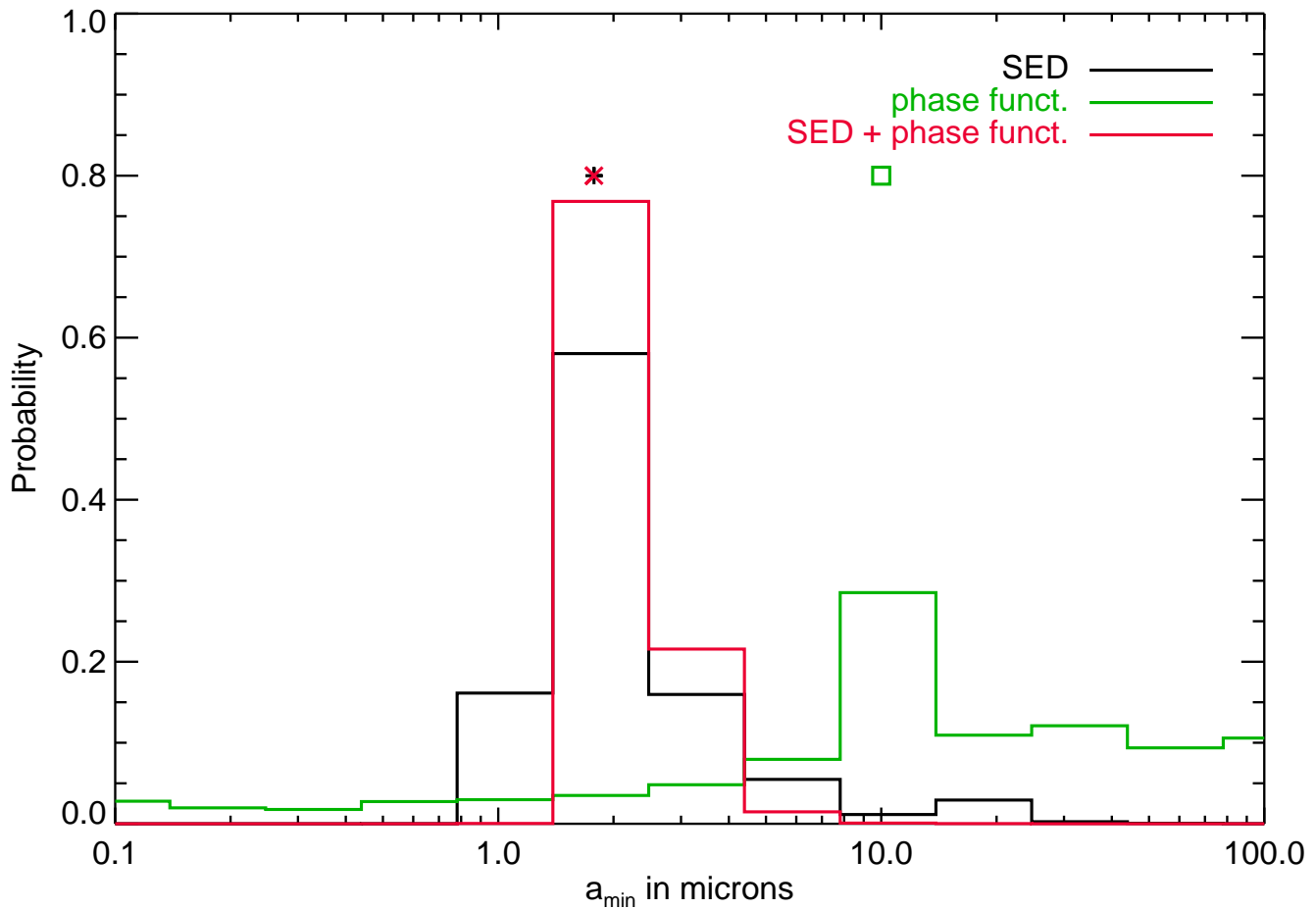}
      \caption{Marginal distributions of the parameter $a_\text{min}$, based on fitting the SED (black histogram), the phase function (green histogram), or both (red histogram). The symbols (crosses and squares) indicate the value of the best models. These distributions are derived from models created using the DHS theory.}
       \label{fig_marginal_PDF_phase}
   \end{figure}

In the scenario $H_1$ (left image),  the synthetic unpolarised images reproduce well the observations in the ansae, with the backward side locally brighter than the forward side. A very strong forward scattering peak is still present on the semi-minor axis, but the reality of this peak cannot be tested in our NICMOS image because we are blind to this separation range. It is however not seen in the NaCo \Lp{} image, although it is predicted in the \Lp{} synthetic scattered light image. We emphasise however that artefacts from the ADI data reduction bias our view of the semi-minor axis, therefore we cannot rule out such a  behaviour. The properties of the grains producing this behaviour are detailed in Table \ref{tab_chi2_Mie_DHS}. These are large \unit{10}{\micro\meter} silicate grains with a power exponent $s=-5.5$. Only DHS models are able to reproduce this interesting behaviour seen in our image. This outcome should be emphasised because it points in the right direction of improved and more realistic dust scattering models. This model provides  a poor fit to the SED with $\chi^2_\text{SED}=32.9 $, however. We marginalised the probability distribution function with respect to the parameter $a_\text{min}$ to illustrate this mismatch (Fig. \ref{fig_marginal_PDF_phase}). The phase function suggests large grains beyond \unit{10}{\micro\meter} to limit the range of the forward-scattering peak to the first $50^\circ$, whereas the SED favours grains below \unit{4}{\micro\meter}. This interesting feature makes sense in the case of aggregate particles. \citet{Volten2007} indeed   experimentally showed that the size of an aggregate as a whole is the dominant factor determining the phase function. These authors further showed that the size of the individual grains making the aggregate is the determining factor for the polarised fraction. Given that the best models for the polarized fraction require $\sim$ \unit{1}{\micro\meter} grains (cf Table \ref{tab_chi2_Mie_DHS}), these experimental results tend to show that the scatterers are  $\sim$ \unit{10}{\micro\meter} aggregates made of $\sim$ \unit{1}{\micro\meter} elementary grains. This scenario is compatible with the SED (red histogram in Fig. \ref{fig_marginal_PDF_phase}) if the size of the elementary grains is also the determining factor for the thermal emissivity of the dust.

In the scenario $H_2$, the best model reproducing the inversion in brightness is much less convincing (Fig. \ref{fig_scattered_light_best_phase_model} right) because the polarised fraction cancels at large scattering angles resulting in a very faint NW side.

\subsubsection{Decrease in the anisotropy of scattering with wavelength}
\label{sec_anistotrpy_scat_vs_lambda}

The second interesting feature revealed by our new reductions is a decrease in the anisotropy of scattering with increasing wavelength, as shown in Fig. \ref{fig_g}. To see if our models can explain this behaviour, we measured the brightness ratio between the NW and SE side of the ring on both our newly reduced Ks and L’ images and our models. The area used to compute this ratio is an elliptical annulus with a semi-major and semi-minor axis of $1.07\arcsec$ and $0.27\arcsec$ and a width of $0.2\arcsec$. As already explained, because of the residual noise affecting the NICMOS Ks image, we only consider the regions in this annulus within $15^\circ$ of the ansa. In Table \ref{tab_results_anisotropy}, we compare these measurements to the predictions of our best SED, scattered light, and overall models. Under the scenario $H_1$, none of them have a SE side brighter than the NW. This was already illustrated in Figure \ref{fig_prediction_I_pI_best_models}. More interestingly here, the brightness asymmetry becomes higher in the \Lp{} band, with a NW:SE ratio further away from unity. This contradicts our measurements pointing towards an isotropic disc at \Lp. Considering the alternative hypothesis $H_2$, here again the NW:SE brightness ratio is far away from unity, especially at \Lp.

\begin{table}
\caption{Comparison between the predictions of the best SED, scattered light, and overall models, and our measurements (in shaded grey) with respect to the NW:SE brightness ratio. The last two rows show the best models that simultaneously fit  the measured NW:SE ratios at Ks and \Lp. The uncertainty is given at a $3\sigma$ level.}             % title of Table
\label{tab_results_anisotropy}      % is used to refer this table in the text
\centering                          % used for centering table
%\begin{tabular}{ c |  p{0.25\linewidth} p{0.25\linewidth}  }        % centered columns (4 columns)
\begin{tabular}{ c |  c c  }        % centered columns (4 columns)
\hline\hline                 % inserts double horizontal lines
Model & NW:SE ratio & NW:SE ratio   \\    % table heading 
 & (Ks) &  (\Lp)   \\    % table heading 
\hline                        % inserts single horizontal line
\rowcolor{Gray}
Measurements &  $0.75\pm0.19$ & $0.95\pm0.14$  \\
\hline                        % inserts single horizontal line
Best $\chi^2_\text{SED}$  & 1.8 & 3.6 \\
Best $\chi^2_{\text{scat. light},H_1}$  & 2.1 & 3.3   \\
Best $\chi^2_{\text{scat. light},H_2}$ & 0.51 & 0.27  \\
 $\chi^2_{\text{overall},H_1}$  & 1.9 & 3.1  \\
 $\chi^2_{\text{overall},H_2}$  & 0.19 & 0.19  \\
Best NW:SE ratio ($H_1$)& 0.82 & 1.25 \\
Best NW:SE ratio ($H_2$)& 0.80 & 0.92 \\
\end{tabular}
\end{table}

\begin{figure}
   \centering
   \includegraphics[width=\hsize]{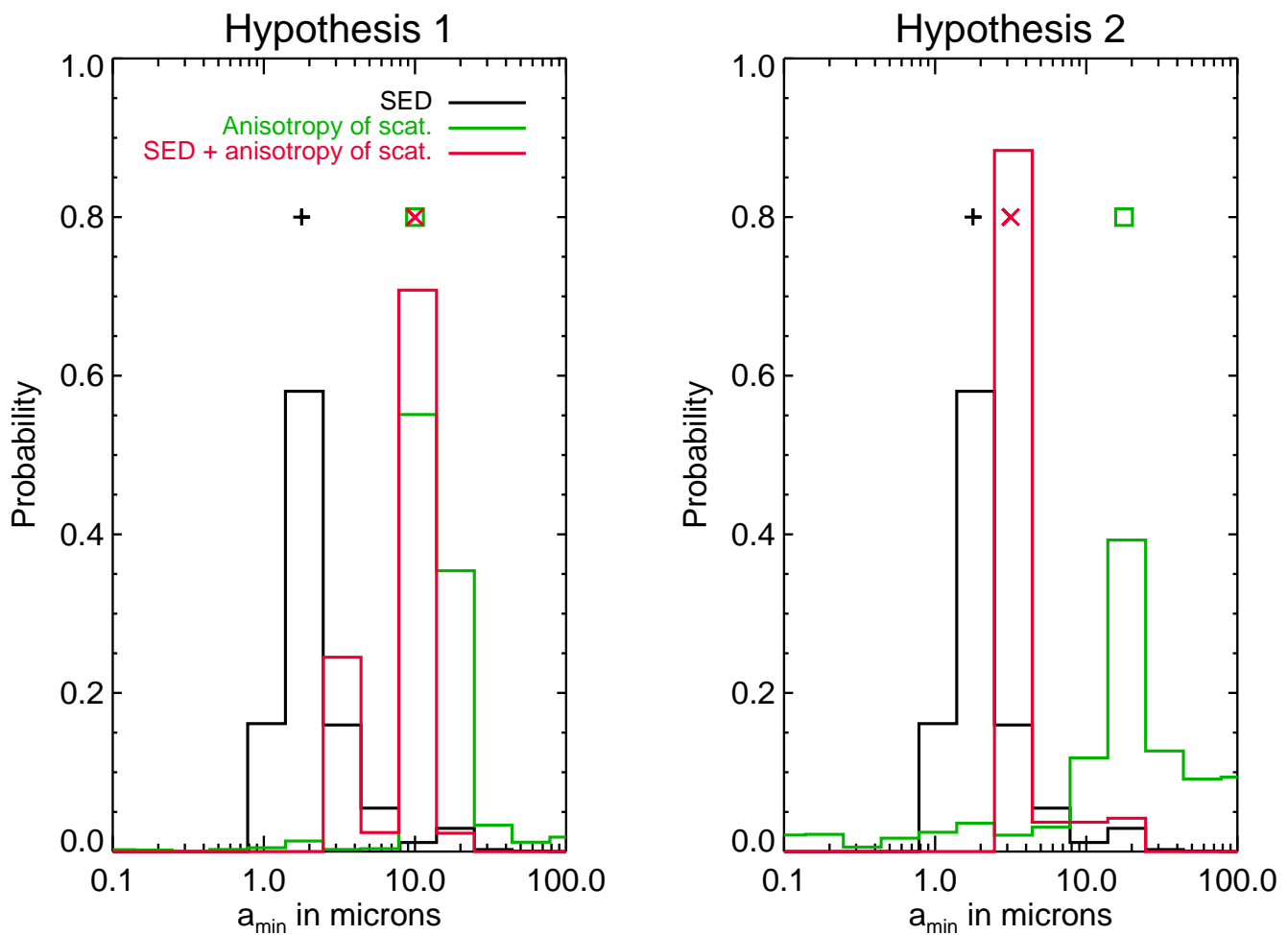}
      \caption{Marginal distributions of the parameter $a_\text{min}$, based on fitting the SED (black histogram), the anisotropy of scattering at Ks and \Lp{} (green histogram), or both (red histogram). The symbols (crosses and squares) indicate the value of the best models. These distributions are derived from models created using the DHS theory.}
       \label{fig_marginal_PDF_anisotropy}
   \end{figure}

We therefore search for models that could explain simultaneously the NW:SE brightness ratios at Ks and \Lp{} and build an additional reduced Chi square with these two measurements. The predictions of the best models are shown in the last two rows of Table \ref{tab_results_anisotropy}. 
If we consider the $H_1$ scenario, then the best reduced Chi square is 5 and the model is compatible with the brightness ratio at Ks but not at \Lp. It is obtained for \unit{10}{\micro\meter} compact silicate grains with very little porosity ($P=0.1\%$ and $p_{H_2O}=90\%$) and with a power exponent $s=-5.5$. This model uses the DHS theory, which, here again, seems a better fit to our data than the Mie theory. Using this Chi square to estimate the goodness of fit of our models with respect to the anisotropy of scattering at Ks and \Lp, we derived the marginal probability distribution with respect to the parameter $a_\text{min}$ (Fig. \ref{fig_marginal_PDF_anisotropy}, left, green histogram). This behaviour can only be explained with grains larger than \unit{10}{\micro\meter}. The best models for the anisotropy of scattering at Ks and \Lp{} are also the best models for the phase function (see Fig. \ref{fig_marginal_PDF_phase}), suggesting that if our assumption of aggregate particles is valid, the overall size of the aggregate is the determining factor for the anisotropy of scattering, rather than the size of the elementary grains. For comparison we overplotted in Fig. \ref{fig_marginal_PDF_phase} the marginal probability distribution based on the SED, favouring \unit{1.8}{\micro\meter} grains. 

In the $H_2$ scenario, the best model has a reduced Chi square of 0.1, and is therefore much more likely. It is obtained with \unit{18}{\micro\metre} pure carbonaceous grains, also with very little porosity ($P=0.1\%$ and $p_{H_2O}=90\%$), and with $s=-4.5$. The marginal probability distribution is shown in Fig. \ref{fig_marginal_PDF_anisotropy} (right) and also poorly agrees with the constraints from the SED, favouring very large grains.

\section{Discussion}
\label{sec_discussion}

Our most striking findings concern the inconsistency between the NW/SE asymmetries in polarised and unpolarised light. We have shown that no matter what is assumed for the side of the disc inclined towards the Earth, no model compatible with the SED can fully explain the scattered light observations. We will now discuss the implications of each assumption regarding the forward-scattering side and propose future observations to answer this question.
\begin{table}   
\caption{Posterior probability of $H_1$ (the NW is inclined towards the Earth) and $H_2$ given  different observables, using the Mie theory. The shaded row highlights the overall conclusion.}             % title of Table
\label{tab_posterior_proba_Mie}      % is used to refer this table in the text
\centering                          % used for centering table
\begin{tabular}{ c | c c  }        % centered columns (4 columns)
\hline\hline                 % inserts double horizontal lines
Observable & $\Psi(H_1|observable)$& $ \Psi(H_2|observable)$ \\    % table heading 
\hline                        % inserts single horizontal line
$p$ & 28.1\% & \textbf{71.9\%} \\
$p\times\phi$ & \textbf{>99.9\% }& <0.1\% \\
$\phi$ & 16.9\% & \textbf{83.1\%} \\
scat. light\tablefootmark{a} & \textbf{>99.9\% }& <0.1\% \\
\rowcolor{Gray}
overall\tablefootmark{b} &\textbf{>99.9\% }& <0.1\% \\
\end{tabular}
\tablefoot{
\tablefoottext{a}{Includes the following independent scattered light observables: polarised fraction $p$, polarised phase function $p\times\phi$, effective albedo and colour index.}
\tablefoottext{b}{Includes the SED and the previous scattered light observables.}
}
\end{table}

\subsection{Scenario $H_1$: The NW side is inclined towards the Earth}

This is the baseline scenario used in our grid of models and it is illustrated by the images in the two left columns of  Fig. \ref{fig_prediction_I_pI_best_models}. The Bayesian analysis shows us that it is the most likely scenario: we computed  the posterior probability of $H_1$ in Table \ref{tab_posterior_proba_Mie} given the different goodness of fit estimators presented in section \ref{sec_estimators}. Although the polarised fraction and the phase function, taken individually,  suggest that the scenario $H_2$ is  more plausible, our new polarised light image strongly supports the opposite conclusion. All in all, the probability of $H_1$ is above 99.9\% given all independent scattered light observations. Further taking the SED into consideration brings the same conclusion. This should not hide the fact that there are still some severe contradictions between observables in this scenario, as noted for instance in section \ref{sec_brightness_inversion}.
The major difficulty of those models is to explain the preferential backward scattering of the grains in unpolarised light. This case is very similar to that of the Fomalhaut debris disc, inclined by about $65^\circ$. The bright side of the ring was shown to be inclined away from us through spectrally resolved interferometric observations \citep{LeBouquin2009}. The best interpretation so far implies very large dust grains of \unit{100}{\micro\metre} whose diffraction peak is confined within a narrow range of scattering angles undetected from the Earth given the system inclination \citep{Min2012}. In the case of HR\ 4796A, this range of scattering angles could be as large as $40^\circ$ since our constraints are very poor in this range in unpolarised light. In these conditions, we have shown in section \ref{sec_brightness_inversion} that \unit{10}{\micro\metre} grains would be sufficient to explain the preferential back-scattering behaviour locally within $15^\circ$ of the ansa (Fig. \ref{fig_scattered_light_best_phase_model} left), at the price of being poorly compatible with the SED. Those grains are also the best candidates to explain the dependance of the anisotropy of scattering with wavelength.
Images at higher angular resolution revealing the disc along its semi-minor axis are clearly required to validate this assumption. 

\subsection{Scenario $H_2$: The NW side is inclined towards the Earth}

Under this assumption, the disc is mainly forward-scattering up to \unit{2.2}{\micro\meter} and the polarised image represents back-scattered light. All our models are preferentially forward-scattering in unpolarised light at all wavelengths between 0.5 and \unit{3.8}{\micro\meter} , which supports this hypothesis. However, this now contradicts the measurements and upper limits on the polarised phase function, which is a very strong argument, as shown in Table \ref{tab_posterior_proba_Mie}.
In this scenario, the offset of the disc detected along the minor axis both in the Ks polarised image and the \Lp{} unpolarised image, could slightly compensate this large asymmetry since the NW side would be closer to the star. Because the flux of the disc scattered light is inversely proportional to the square of the distance from the star, if we assume an offset of 5.5AU, the NW side would only be 35\% brighter, which is still not sufficient to make the polarised light image compatible with our measurements.

\subsection{Validation of the optically thin hypothesis}
\label{sec_optical_thickness}
All models presented here rely on the assumptions that the disc is optically thin. We validated this hypothesis by re-running the simulations of the best models with the correct disc mass necessary to fit the SED. The best overall model under hypothesis $H_1$ predicts a disc mass of $0.5$\Mearth{} distributed among particles between \unit{1}{\micro\metre} and 10mm. The opacity reaches 0.08 perpendicular to the midplane of the disc and 1.0 along the mid-plane. These mass-corrected models differ from the optically-thin models by a few percent, therefore increasing the Chi squares by the same amount. Therefore we cannot exclude that taking this effect into account would change the best parameters presented here by a few percent. The total dust mass of the best SED, scattered light and overall models presented here is always below or equal to $0.5$\Mearth, and no larger deviations from the optically thin models are expected. We investigated in a few examples whether a more massive disc could explain the change in the brightest side with wavelength, but could not find models compatible with our observations. An idea proposed to explain the spectral behaviour of the dust is indeed that the disc is optically thick in the optical and up to the Ks band and becomes optically thin in the \Lp{} band \citep{Perrin2014}. However, we could not recreate this change in brightness. In particular, even for an optically thick disc, the brightest side remained the forward scattering side. We noticed from these models that the size of the ring should appear smaller for an optically thick disc, because most of the light is emitted by the inner regions of the ring. We do not notice this change within error bars, however. This analysis goes beyond the initial scope of our paper and we leave the analysis of the transition between an optically thin and an optically thick disc to a further study.

\subsection{Follow-up observations}

The questions raised by this new polarimetric view of the disc clearly call for further work on the observational and theoretical sides. A limitation of the current observations in unpolarised light is the inability to reveal the disc along the minor axis. This is however of great interest because the models predict the highest contrast between the forward and backward scattering side at this location. Thanks to
the improved PSF stability, these observations are now possible with instruments such as SPHERE \citep{Beuzit2008} or GPI \citep{Macintosh2014}, using reference star subtraction instead of ADI to avoid the problem of disc self-subtraction at such a short separation \citep{Milli2012}. With their polarimetric capabilities, a determination of the scattering phase function and polarised fraction at all angles scattered towards the Earth will be possible. With the near-infrared instrument SPHERE/IRDIS \citep{Langlois2010}, we can expect to reveal the spectral dependance of the phase matrix from the K band to the Y band, and maybe down to the V band with the visible instrument SPHERE/ZIMPOL \citep{Schmid2010} with sufficient observing time. These observations would be of great interest to compare the empirical phase functions and polarised fractions to that measured in laboratory on cosmic dust analogues (see for instance \citealt{Volten2007}). This could additionally validate our assumption that the dust is made of $\sim$ \unit{10}{\micro\meter} aggregates made of $\sim$ \unit{1}{\micro\meter} elementary grains. Resolved images of the thermal emission of the dust with the radiotelescope ALMA should provide additional constraints on the dust mass and thus the optical thickness of the disc in scattered light, as  was already done for Fomalhaut \citep{Boley2012} and $\beta$ Pictoris \citep{Dent2014}.

\section{Conclusions}

Our new observations show a clear detection of the polarised light of the debris disc surrounding HR\ 4796A. Thanks to the PDI technique, the disc morphology can be probed more accurately at short separations, close to the minor axis.
Starting from constraints on the dust grains based on previous modelling, we explore the compatibility of a grid of models with the scattered light both in intensity and polarisation as predicted by two theories of light scattering: EMT-Mie and EMT-DHS. Our results confirm the earlier modelling suggesting that grains larger than \unit{1}{\micro\meter} are needed to explain the mid- to far-infrared excess of the star. 
Both theories predict a strong forward/backward brightness asymmetry in polarised and unpolarised light, which is not consistent with the observational constraints showing that the SE side is brighter in unpolarised light from the visible up to \unit{2.2}{\micro\meter}, whereas the NW side is brighter in polarisation. We explore different scenarios to explain this apparent contradiction. Two conclusions emerge from this work. First, this shows that the dust particles are probably not spherical, as already pointed out by \citet{Debes2008} and \citet{Augereau1999}, but made of aggregates of micronic particles with two distinct behaviours, whether we consider the thermal properties, the unpolarised phase function, or the polarised phase function. This is why the Mie theory and, to a smaller extent, the statistical approach of the DHS theory are probably unadapted to reproduce the scattering properties of these irregular fluffy aggregates. Even if not perfect, we note that the DHS theory is going in the right direction and  provides  a closer match to the scattered light images than the Mie theory. Secondly, the case of HR\ 4796A is probably very similar to that of Fomalhaut where backward-scattering was already shown to challenge these traditional theories of scattering by spherical particles. A new generation of models is clearly needed, and observational constraints will come along to refine them. In the optical and near-infrared, high-resolution imagers such as GPI and SPHERE will be able to test whether a diffraction peak is observed for small scattering angles, as predicted by the Mie theory for circular grains, and will measure with high accuracy the phase function and polarised fraction over a much wider range of scattering angles and at various wavelengths.

\begin{acknowledgements}
J.M. acknowledges financial support from the ESO studentship program and the Labex OSUG2020. C. Pinte acknowledges funding from the Agence Nationale pour la Recherche (ANR) of France under contract ANR-2010-JCJC-0504-01.
We would like to thank John Krist for his initial help with the NICMOS data and we thank ESO staff and technical operators at the Paranal Observatory.
\end{acknowledgements}

\bibliography{biblio_HR4796}    

\Online
\begin{appendix} %First online appendix

\section{Marginal distributions of the free parameters using the DHS theory}
\label{App_marginal_PDF_DHS}
   
\begin{figure*}[b]
   \centering
  \includegraphics[width=\hsize]{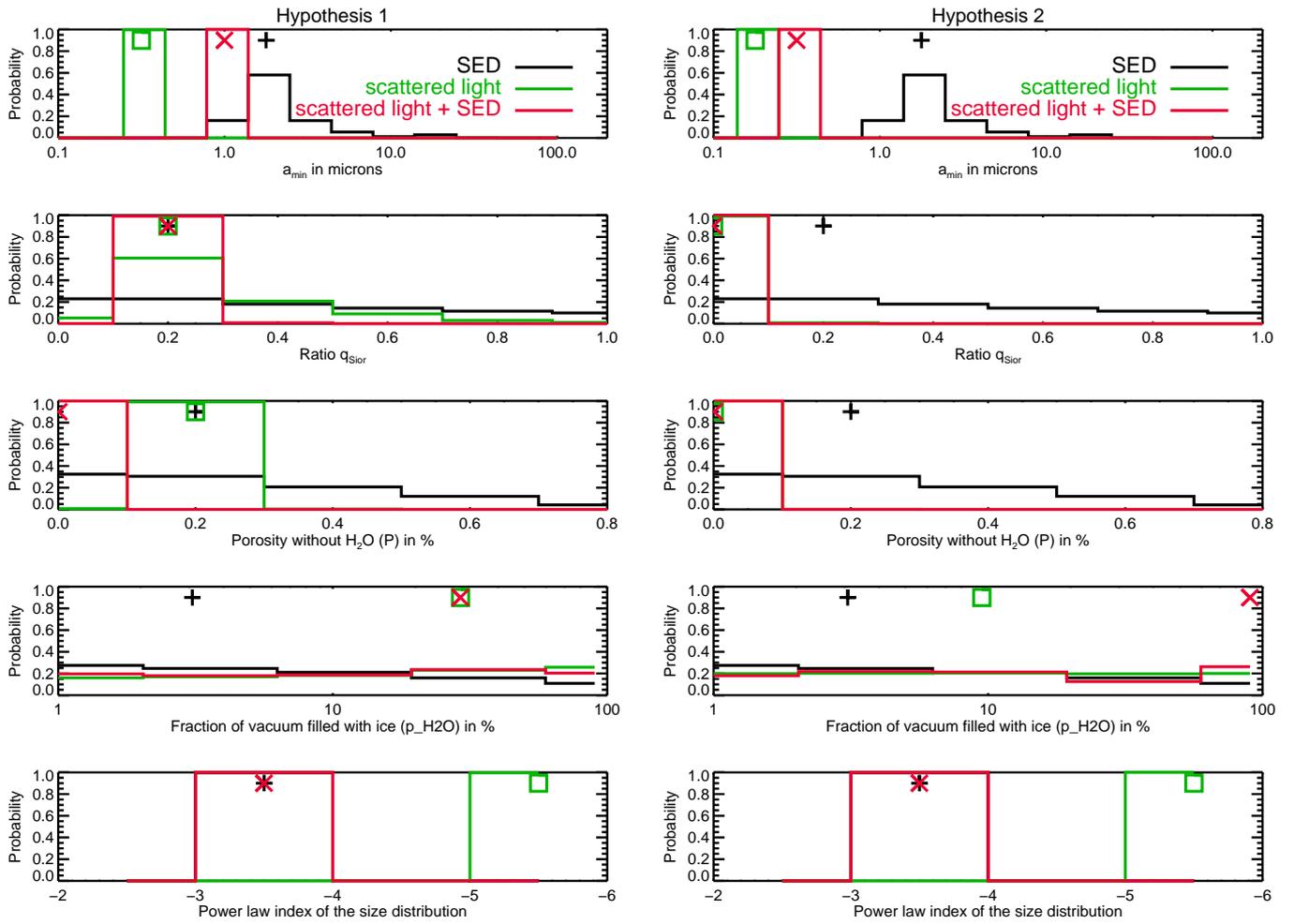}
   \caption{Marginal distributions of the five free parameters of our model, based on fitting the SED (black histogram), the scattered light images (polarised fraction, polarised phase function, colour index and effective albedo, green histogram) or both (red histogram). The symbols (crosses and squares) indicate the value of the best SED, scattered light or overall model. These distributions are derived from models created using the DHS theory.}
       \label{fig_marginal_PDF_DHS}
   \end{figure*}

   We show in Fig. \ref{fig_marginal_PDF_DHS} the marginal distributions of the five free parameters $(a_{min},q_{Sior},P,p_{H_2O},s)$ obtained with the DHS theory. Using this theory, the scattered light appears incompatible with the SED, regarding the parameters $a_{min}$ and $s$, in both scenarios $H_1$ and $H_2$. Regarding the parameters $q_{Sior}$, $P$, and $p_{H_2O}$, the SED does not provide constraints strong enough to draw significant conclusions.
   
\end{appendix}

\end{document}